\documentclass[twocolumn, trackchanges]{aastex63}

\newcommand{\rthree}{FRB\,20180916B}
\newcommand{\rthreefull}{FRB\,20180916B}
\newcommand{\rthreehost}{SDSS\,J015800.28+654253.0}
\newcommand{\rone}{FRB\,20121102A}
\newcommand{\sgr}{SGR\,1935+2154}
\newcommand{\halpha}{\ensuremath{\mathrm{H\alpha}}}

\newcommand{\msun}{\ensuremath{\mathrm{M_\odot}}}
\newcommand{\hst}{\emph{HST}}
\usepackage{color}
\usepackage{amsfonts}
\usepackage{amsmath}
\usepackage{amssymb}

\submitjournal{\apjl}

\shorttitle{Environment of FRB~20180916B}
\shortauthors{Tendulkar et al.}

\begin{document}

\title{The 60-pc Environment of \rthreefull}

\author[0000-0003-2548-2926]{Shriharsh P. Tendulkar}
\affiliation{Department of Astronomy and Astrophysics, Tata Institute of Fundamental Research, Homi Bhabha Road, Colaba, Mumbai, Maharashtra, 400005, India}
\affiliation{National Centre for Radio Astrophysics, Pune University Campus, Post Bag 3, Ganeshkhind, Pune, Maharashtra, 411007, India}

\author[0000-0001-6150-2854]{Armando Gil de Paz}
\affiliation{Departamento de Física de la Tierra y Astrofísica, Universidad Complutense de Madrid (UCM), Plaza Ciencias 1, Madrid, E-28040, Spain}
\affiliation{Instituto de Física de Partículas y del Cosmos (IPARCOS), Plaza Ciencias 1, Madrid, E-28040, Spain} 

\author[0000-0002-8139-8414]{Aida Yu. Kirichenko}
\affiliation{Instituto de Astronomía, Universidad Nacional Autónoma de M\'exico, Apdo. Postal 877, Ensenada, Baja California 22800, M\'exico}
\affiliation{Ioffe Institute, 26 Politekhnicheskaya st., St. Petersburg 194021, Russia}

\author[0000-0003-2317-1446]{Jason~W.~T.~Hessels}
\affiliation{Anton Pannekoek Institute for Astronomy, University of Amsterdam, Science Park 904, 1098 XH, Amsterdam, The Netherlands}
\affiliation{ASTRON, Netherlands Institute for Radio Astronomy, Oude Hoogeveensedijk 4, 7991 PD Dwingeloo, The Netherlands}

\author[0000-0002-3615-3514]{Mohit Bhardwaj}
\affiliation{Department of Physics, McGill University, 3600 rue University, Montr\'eal, QC H3A 2T8, Canada}
\affiliation{McGill Space Institute, McGill University, 3550 rue University, Montr\'eal, QC H3A 2A7, Canada} 


\author{Fernando \'Avila}
\affiliation{Observatorio Astron\'omico Nacional, Instituto de Astronom\'ia, Universidad Nacional Aut\'onoma de M\'exico, Ensenada, Baja California, M\'exico}

\author[0000-0002-1429-9010]{Cees Bassa}
\affiliation{ASTRON, Netherlands Institute for Radio Astronomy, Oude Hoogeveensedijk 4, 7991 PD Dwingeloo, The Netherlands}

\author[0000-0002-3426-7606]{Pragya Chawla}
\affiliation{Department of Physics, McGill University, 3600 rue University, Montr\'eal, QC H3A 2T8, Canada}
\affiliation{McGill Space Institute, McGill University, 3550 rue University, Montr\'eal, QC H3A 2A7, Canada} 

\author[0000-0001-8384-5049]{Emmanuel Fonseca}
\affiliation{Department of Physics, McGill University, 3600 rue University, Montr\'eal, QC H3A 2T8, Canada}
\affiliation{McGill Space Institute, McGill University, 3550 rue University, Montr\'eal, QC H3A 2A7, Canada} 
\affiliation{Department of Physics and Astronomy, West Virginia University, P.O. Box 6315, Morgantown, WV 26506, USA}
\affiliation{Center for Gravitational Waves and Cosmology, West Virginia University, Chestnut Ridge Research Building, Morgantown, WV 26505, USA}

\author[0000-0001-9345-0307]{Victoria M. Kaspi}
\affiliation{Department of Physics, McGill University, 3600 rue University, Montr\'eal, QC H3A 2T8, Canada}
\affiliation{McGill Space Institute, McGill University, 3550 rue University, Montr\'eal, QC H3A 2A7, Canada} 

\author[0000-0002-5575-2774]{Aard Keimpema}
\affiliation{Joint Institute for VLBI ERIC, Oude Hoogeveensedijk 4, 7991~PD Dwingeloo, The Netherlands}

\author[0000-0001-6664-8668]{Franz Kirsten}
\affiliation{Department of Space, Earth and Environment, Chalmers University of Technology, Onsala Space Observatory, 439 92, Onsala, Sweden}

\author{T. Joseph W. Lazio}
\affiliation{Jet Propulsion Laboratory, California Institute of Technology, 4800 Oak Grove Dr, Pasadena, CA 91109, \hbox{USA}}

\author[0000-0001-9814-2354]{Benito Marcote}
\affiliation{Joint Institute for VLBI ERIC, Oude Hoogeveensedijk 4, 7991~PD Dwingeloo, The Netherlands}

\author[0000-0002-4279-6946]{Kiyoshi Masui}
\affiliation{MIT Kavli Institute for Astrophysics and Space Research, Massachusetts Institute of Technology, 77 Massachusetts Ave, Cambridge, MA 02139, USA}
\affiliation{Department of Physics, Massachusetts Institute of Technology, 77 Massachusetts Ave, Cambridge, MA 02139, USA}

\author[0000-0003-0510-0740]{Kenzie Nimmo}
\affiliation{ASTRON, Netherlands Institute for Radio Astronomy, Oude Hoogeveensedijk 4, 7991 PD Dwingeloo, The Netherlands}
\affiliation{Anton Pannekoek Institute for Astronomy, University of Amsterdam, Science Park 904, 1098 XH, Amsterdam, The Netherlands}

\author[0000-0002-5195-335X]{Zsolt Paragi}
\affiliation{Joint Institute for VLBI ERIC, Oude Hoogeveensedijk 4, 7991~PD Dwingeloo, The Netherlands}

\author[0000-0003-1842-6096]{Mubdi Rahman}
\affiliation{Dunlap Institute for Astronomy \& Astrophysics, University of Toronto, 50 St. George Street, Toronto, ON M5S 3H4, Canada}
\affiliation{Sidrat Research, Toronto, ON, Canada}

\author[0000-0003-0456-149X]{Daniel Reverte Pay\'a}
\affiliation{Instituto de Astrof\'isica de Canarias, V\'ia L\'actea s/n, E38200, La Laguna, Tenerife, Spain}
\affiliation{GRANTECAN, Cuesta de San Jos\'e s/n, E-38712, Bre\~{n}a Baja, La Palma, Spain}

\author[0000-0002-7374-7119]{Paul Scholz}
\affiliation{Dunlap Institute for Astronomy \& Astrophysics, University of Toronto, 50 St. George Street, Toronto, ON M5S 3H4, Canada}

\author[0000-0001-9784-8670]{Ingrid Stairs}
\affiliation{Department of Physics \& Astronomy, University of British Columbia, 6224 Agricultural Road, Vancouver, BC V6T 1Z1, Canada}

\correspondingauthor{Shriharsh P. Tendulkar}
\email{shriharsh.tendulkar@tifr.res.in}

\begin{abstract}
Fast Radio Burst \rthree\ in its host galaxy \rthreehost\ at 149\,Mpc is by far the closest-known FRB with a robust host galaxy association. The source also exhibits a 16.35-day period in its bursting. Here we present optical and infrared imaging as well as integral field spectroscopy observations of \rthreefull\ with the WFC3 camera on the \emph{Hubble Space Telescope} and the MEGARA spectrograph on the 10.4-m Gran Telescopio Canarias. The 60--90 milliarcsecond (mas) resolution of the \emph{Hubble} imaging, along with the previous 2.3-mas localization of \rthreefull, allow us to probe its environment with a 30--60\,pc resolution. We constrain any point-like star-formation or \ion{H}{2} region at the location of \rthreefull\ to have an \halpha\ luminosity $L_\mathrm{H\alpha} \lesssim 10^{37}\,\mathrm{erg\,s^{-1}}$ and, correspondingly, constrain the local star-formation rate to be $\lesssim10^{-4}\,\mathrm{\msun\,yr^{-1}}$. The constraint on \halpha\ suggests that possible stellar companions to \rthreefull\ should be of a cooler, less massive spectral type than O6V. \rthreefull\ is 250\,pc away (in projected distance) from the brightest pixel of the nearest young stellar clump, which is $\sim380$\,pc in size (full-width at half maximum). With the typical projected velocities of pulsars, magnetars, or neutron stars in binaries ($60-750$\,km~s$^{-1}$), \rthreefull\ would need 800\,kyr to 7\,Myr to traverse the observed distance from its presumed birth site. This timescale is inconsistent with the active ages of magnetars ($\lesssim10$\,kyr). Rather, the inferred age and observed separation are compatible with the ages of high-mass X-ray binaries and gamma-ray binaries, and their separations from the nearest OB associations.
\end{abstract}


\section{Introduction} 
\label{sec:intro}
More than a decade after the discovery of the `Lorimer Burst' \citep{lbm+07}, the physical origin(s) of Fast Radio Bursts (FRBs) remains unclear.  These bright ($\sim$\,0.1--100\,Jy~ms fluence), short-duration ($\sim$\,{\textmu}s$-100$\,ms) radio flashes have been shown to be extragalactic in origin, but it is still unclear what type of object produces them, or what the exact emission mechanism might be --- see \citet{phl19} and \citet{cc19} for recent reviews and \citet{pww+18} for a catalogue of proposed theories. The high brightness temperatures ($T_b \sim 10^{36}$\,K) of FRBs point to coherent emission from a compact source with high energy density, and for this reason many models have invoked neutron stars, white dwarfs and/or black holes in a variety of possible settings. The fact that some FRB sources are repeating \citep{ssh+16a}, whereas others appear to be one-off events \citep{smb+18}, also raises the question of whether the phenomenon can be ascribed to a single source type, or whether there are at least two sub-populations with distinct physical natures \citep{cui2021}.

Detailed spectro-temporal and polarimetric characterisation of FRB signals can help constrain models \citep{ffb18,day2020,nimmo2020}, as can multi-wavelength associations or constraints \citep{sbh+17,bhandari2020, scholz2020}.  Radio interferometers have now provided robust host galaxy associations for roughly a dozen FRBs\footnote{\url{http://frbhosts.org/}} --- both repeating and apparently one-off \citep{clw+17,rcd+19,bdp+19,pmm+19,mnh+20,mpm+20}.  In principle, the properties of these host galaxies also offer important clues, but thus far a wide range of host galaxy types have been observed \citep{bsp+20,heintz2020, mannings2020}.  FRB models need to accommodate this diversity or resort to multiple populations.  High-precision ($\lesssim 100$\,mas) positions are possible with the Australia Square Kilometre Array Pathfinder \citep[ASKAP;][]{bdp+19}, Very Large Array \citep[VLA;][]{clw+17} and European Very-long-baseline-interferometry Network \citep[EVN;][]{mph+17}, and allow for localisation {\it within} host galaxies.  This can, e.g., confirm or exclude an association with the nucleus of the host galaxy, or a region of active star-formation.

Insights into the FRB mystery can also come from finding analogous sources in our own Milky Way.  The recent discovery of an exceptionally bright ($\sim$MJy~ms) radio burst \citep[sometimes designated FRB\,200428A;][]{abb+20,brb+20} --- and accompanying hard X-ray burst \citep{mereghetti2020, ridnaia2020,tavani2020, li2020} --- from the Galactic magnetar \sgr\ bridges the many orders-of-magnitude in luminosity between the pulses (regular and giant) of canonical radio pulsars and the extragalactic FRBs.  In fact, if placed at the 149-Mpc distance of the closest localised FRB, \sgr's bright burst would only be a factor of $\sim30$ times less luminous compared to the least energetic FRBs seen thus far.  This suggests that a significant fraction of FRBs could have a magnetar origin.  The discovery of 20--100\,Jy\,ms bursts from \sgr\ \citep{kirsten2020} --- bright compared to FRB fluences but far weaker in luminosity --- might also suggest that we are only seeing the tip of the burst energy distribution from extragalactic FRBs.

\rone\ \citep[previously FRB 121102;][]{sch+14,ssh+16a} and \rthreefull\ \citep[previously FRB 180916.J0158+65;][]{abb+19c} are currently the two best-characterised repeating FRBs, and the first two to be precisely localised within a host galaxy \citep{clw+17,tbc+17,mph+17,mnh+20}. The spectro-temporal and intrinsic polarimetric properties of bursts from \rone\ and \rthree\ are remarkably similar, strongly suggesting that they have the same progenitor type and detailed emission mechanism.  

Both sources show the characteristic downward frequency drift between sub-bursts \citep{hss+19}, which is seen in bursts from many repeaters \citep{abb+19b, abb+19c, fab+20} and colloquially termed the `sad trombone' effect. \rone\ showed a $30$-{\textmu}s-wide burst component at 4.5\,GHz \citep{msh+18}; a recent study of \rthree\ using voltage data finds burst structure down to $\sim$ 3--4\,{\textmu}s, and spanning close to 3 orders-of-magnitude up to $\sim2$\,ms within individual bursts \citep{nimmo2020}.  

These two repeaters also have indistinguishable polarimetric properties, showing nearly 0\% circular polarization, but $\sim 100\%$ linear polarization with a roughly flat polarization angle during and {\it between} bursts\footnote{However, another repeating source, FRB 20180301A (previously FRB\,180301), was recently shown to have diverse polarization swings and polarization fractions in different bursts \citep{luo2020} showing that \rthree\ and \rone\ are not necessarily representative of the whole repeater population.} \citep{msh+18,nimmo2020}.  However, \rone\ shows a drastically higher Faraday rotation measure \citep[RM$\sim 10^5\,\mathrm{rad\,m}$;][]{msh+18} which is highly variable ($\Delta\mathrm{RM}\sim 3\times10^4\,\mathrm{rad\,m^{-2}}$) on timescales of days to years \citep{msh+18,gsp+18,hilmarsson2020b}.  \rone\ also shows clear dispersion measure (DM) variations \citep[$\Delta\mathrm{DM} \sim 3-5\,\mathrm{pc\,cm^{-3}}$;][]{hss+19,jcf+19}.  In comparison, \rthree\ shows a much more stable DM \citep[$\Delta\mathrm{DM}\lesssim 0.1\,\mathrm{pc~cm^{-3}}$;][]{r3_period_2020,nimmo2020}, and RM variations of only $\sim 2-3\,\mathrm{rad~m^{-2}}$ \citep{pleunis2020}.

It has recently also been reported that both sources are periodic in their activity, with \rthree\ modulated at $P_{\rm activity} = 16.33\pm0.12$\,day \citep{r3_period_2020,pleunis2020} and \rone\ likely modulated at $P_{\rm activity} \sim 160$\,day \citep{rms+20, cruces2020}.  This could conceivably be related to an orbital period \citep{iokazhang2020, lyutikov2020, zhanggao2020, popov2020}, rotational period \citep{beniamini2020}, or precession period \citep{levin2020, sobyanin2020, yangzou2020, zanazzi2020}. 

At first glance, perhaps the most striking difference between these astrophysical sources is their host galaxy and local environment: \rone\ is localised to a low-metallicity dwarf at $z = 0.193$ \citep{clw+17,tbc+17}, while \rthree\ is found in a massive $10^{10}\,\msun$ spiral at $z = 0.0337$ \citep{mnh+20}. Both sources are found in close proximity to a prominent star-forming region \citep{bassa2017frb,mnh+20}, though \rone's $\sim10$ milliarcsecond (mas) localisation \citep{mph+17}, {\em Hubble Space Telescope} (\hst) imaging \citep{bassa2017frb}, and adaptive optics observations \citep{kokubo2017} demonstrate that it is offset by $\sim 200$\,pc from the peak of star-formation in this region\footnote{Yet another repeater, FRB 20190711 \citep{macquart2020} is also found in a massive $8\times10^{9}\,\mathrm{\msun}$ star-forming galaxy \citep{heintz2020} though the localization of the FRB is too imprecise to identify its local environment.}. Lastly, the association of \rone\ with a persistent (isotropic luminosity $\sim 10^{39}$\,erg~s$^{-1}$) and compact ($< 0.7$\,pc) radio source \citep{clw+17,mph+17} stands in stark contrast to the lack of such a counterpart to \rthree\ \citep{mnh+20}, despite it being significantly nearer to Earth. 

At a luminosity distance of 149\,Mpc, \rthree\ is by far the closest-known FRB with a robust host galaxy association \citep{mnh+20}.  It is also the most precisely localised FRB to date: EVN observations achieved a 2.3-mas localisation within the International Celestial Reference Frame (ICRF), including systematic uncertainties \citep{mnh+20}. \rthree\ thus provides an unprecedented opportunity for high-resolution optical studies of its local environment.  In previous Gemini North observations, \rthree\ was associated with the apex of an apparently `V'-shaped star-forming region (or complex of closely spaced star-forming regions) with an extent of $\sim2^{\prime\prime}$.  Given the 0.8--1.0\arcsec\ seeing of those observations, higher-resolution observations can greatly enhance our understanding of \rthree's local environment, and perhaps even detect a massive binary companion that could elucidate its periodic activity.

Here we present an imaging and spectroscopic study of the immediate environment of \rthree\ using observations from \hst\ and the 10.4-m Gran Telescopio Canarias (GTC).  These observations probe 60-pc scales within the host galaxy --- by far the closest view of any FRB source to date.  We present the observations and analysis in \S2 and \S3, respectively, and discuss the astrophysical implications and interpretation in \S4.

\section{Observations \& Reduction}
We observed \rthree\ using the Multi-Espectr\'ografo en GTC de Alta Resolución para Astronomía (MEGARA) integral field spectrograph on the GTC and the Wide Field Camera 3 (WFC3) instrument on \hst\ in 2019 September and 2020 July (Table~\ref{tab:obs}). Here we describe the observations and data reduction procedures.

\subsection{MEGARA}

Observations of \rthreehost\ were performed with the MEGARA instrument \citep{gildepaz2018,carrasco2018} at the 10.4-m GTC in 2019 September (see Table~\ref{tab:obs} for details). We used the Large Compact Bundle Integral Field Unit mode (LCB IFU), which provides a field of view (FoV) of 12\farcs5 $\times$ 11\farcs3 and a spectral pixel (spaxel) size of 0\farcs62 ($\equiv450\,\mathrm{pc}$ at 149\,Mpc). The observations were carried out using the LR-R setup with a spectral coverage of 6100--7300\,\AA. The pointing was set so that the MEGARA FoV covered both the `V'-shaped structure found near \rthree\ as well as the host galaxy nucleus (see Figure~\ref{fig:gtc}, panel {\bf a}). During the run we also observed the spectrophotometric standard star HR7596, and acquired halogen lamp flats and ThNe arcs using the MEGARA Instrument Calibration Module (ICM), as well as a series of bias images. 

The data were processed using the development version (v0.9.2) of the MEGARA Data Reduction Pipeline\footnote{\url{https://github.com/guaix-ucm/megaradrp}}  \citep[DRP;][]{pascual2018,pascual2019}, which is based on a series of processing recipes, and the cookbook\footnote{\url{http://www.gtc.iac.es/instruments/megara/media/MEGARA\_cookbook\_1I.pdf}}. The halogen lamp observations allowed us to trace the spectra (\texttt{TraceMap} recipe), to recover the flux of each fiber affected by cross-talk contamination from adjacent fibers (\texttt{ModelMap} recipe), and to correct for changes in sensitivity from blue-to-red in between fibers (\texttt{FiberFlat} recipe). Prior to the correction by fiber-flat, we wavelength-calibrated the images (including the master fiber-flat), fiber-by-fiber, using ThNe arc observations obtained with the MEGARA ICM. The \texttt{LcbStdStar} recipe allowed us to use the LCB observations of the standard stars to derive the system response function after assuming the La Palma extinction curve\footnote{\url{https://www.ing.iac.es/Astronomy/observing/manuals/ps/tech\_notes/tn031.pdf}}. The results from all these recipes were finally used (\texttt{LcbImage} recipe) to process the \rthreehost\ data. The sky background subtraction was performed using the processed fiber spectra of the 8 fixed 7-fiber minibundles (56 fibers) that are mounted on the LCB pseudo-slit and that are placed in a blank sky region 1\farcm75--2\arcmin\ away from the center of the LCB (which is also the optical axis of the instrument). The final product of this data reduction is a Row-Stacked Spectra (RSS hereafter) FITS file which is wavelength and flux calibrated and has its sky background subtracted. 

\subsection{WFC3}
The host galaxy of \rthree\ was observed with the WFC3 instrument on the \hst\ in the  F657N (6476--6674\AA) and F673N (6681--6880\AA) filters of the UVIS channel, as well as the F110W filter (8832--14121\AA, IR channel) on 2020 July 16 \& 17. Table~\ref{tab:obs} summarizes the observations. The aim of the F110W observations was to detect or constrain the presence of bright stars or stellar clusters and to understand the morphology of the environment. At the redshift z=0.0338 of the host galaxy, the \halpha\ line is shifted to 6784\AA\ (within the F673N filter coverage) while the zero-redshift \halpha\ filter, F657N, is used as an \halpha-off filter to constrain the underlying continuum. The \halpha-on and -off images are acquired to constrain local star-formation and understand the distribution of atomic hydrogen in the region. At 149\,Mpc, the angular and spatial resolution of the F657N, F673N, and F110W filters is 56\,mas$\equiv40\,\mathrm{pc}$, 58\,mas$\equiv42\,\mathrm{pc}$, and 95\,mas$\equiv68\,\mathrm{pc}$, respectively.  

The UVIS observations were undertaken with 3 exposures of 959\,s each (with a total exposure of 2877\,s), dithered in the 3-point dither pattern (\texttt{WFC3-UVIS-DITHERLINE-3PT}) to optimally sample the PSF. A post-exposure flash adjusted for 9 electrons was used to minimze CTE losses, as recommended by the WFC3/UVIS data handbook\footnote{\url{https://hst-docs.stsci.edu/wfc3dhb}}. The IR observations were undertaken with 10 exposures with a 4-point dither pattern (\texttt{WFC3-IR-DITHERBOX-MIN}) read-out with the \texttt{SPARS50} readout sequence, for a total exposure of 5929\,s.

The pre-calibrated and CTE-corrected UVIS and IR (\texttt{.FLC} and \texttt{.FLT}, respectively) images were distortion-corrected and astrometrically aligned to the International Celestial Reference Frame (ICRF) using the \emph{Gaia} DR2 catalog \citep{gaiamission, gaiadr2} and the \texttt{tweakreg} tool. The images were individually aligned using 60--90 stars (UVIS images) and 30--35 stars (IR images) to achieve a typical astrometric root-mean-square (RMS) residual of 18\,mas and 36\,mas in the UVIS and IR images, respectively. The alignment error between the Gaia optical reference frame and the ICRF frame defined with radio sources is negligible in this context.

The aligned images were combined using \texttt{astrodrizzle} to make cosmic-ray rejected images with a final platescale of 30-mas/pixel. Photometry was performed on the aligned, individual exposures \citep[\texttt{dolphot}][]{dolphot2016} using the appropriate point-spread functions for each filter. 

There is no point-source detected at the location of \rthree. To constrain the detectable source brightness, including the underlying diffuse emission from the host galaxy, we used the \texttt{addstars} tool to add simulated point sources at the location of \rthree\ with a range of brightnesses and calculated the detection significance through \texttt{dolphot} in each filter.

We use the absolute photometric calibration as defined by the WFC3 calibration team\footnote{\url{https://www.stsci.edu/hst/instrumentation/wfc3/data-analysis/photometric-calibration}}, which has systematic errors of $\approx2\%$ (F110W) and $\sim10\%$ (F657N, F673N).

\begin{deluxetable*}{clcccl}
\tablecaption{Observation Details\label{tab:obs}}
\tablecolumns{6}
\tablewidth{0pt}
\tablehead{
\colhead{Obs. Date} & \colhead{Instrument/} & \colhead{Grating/} & \colhead{Exp. Time} & \colhead{Obs ID$^\mathrm{a}$} & \colhead{Notes$^\mathrm{b}$} \\
\colhead{(UTC)} & \colhead{Camera} & \colhead{Filter} & \colhead{(s)} & \colhead{} & \colhead{}
}
\startdata
\cutinhead{Gran Telescopio Canarias (Program GTC18-19BMEX)}
2019-09-24 & MEGARA/IFU & LR-R & 680 & 2303712 & AM=1.27, seeing=1\farcs0 \\
2019-09-24 & MEGARA/IFU & LR-R & 680 & 2303713 & AM=1.26, seeing=1\farcs0 \\
2019-09-24 & MEGARA/IFU & LR-R & 680 & 2303714 & AM=1.26, seeing=1\farcs0 \\
\cutinhead{\emph{Hubble Space Telescope} Program (16072)}
2020-07-16 &  WFC3/UVIS1 &  F673N &  2877 & IE8Q01010 & resolution=0\farcs058\\
2020-07-17 &  WFC3/UVIS1 &  F657N &  2877 & IE8Q02010 & resolution=0\farcs056\\
2020-07-17 &  WFC3/IR    &  F110W &  306  & IE8Q03010 & resolution=0\farcs095\\
2020-07-17 &  WFC3/IR    &  F110W &  5623 & IE8Q03020 & resolution=0\farcs095\\
\enddata
\tablecomments{ a: Observation ID for the GTC Archive \url{https://gtc.sdc.cab.inta-csic.es/gtc/jsp/searchform.jsp} and the HST MAST Archive: \url{https://archive.stsci.edu/index.html}. b: Airmass (AM) and seeing conditions and full-width at half maximum of the point spread function for WFC3. The size of the spaxel for MEGARA observations is 0\farcs62.}
\end{deluxetable*}

\section{Analysis \& Results}

\subsection{IFU Spectroscopy}
\begin{figure*}
\begin{minipage}[h]{1.\linewidth}
\center{
\includegraphics[width=0.46\linewidth,clip]{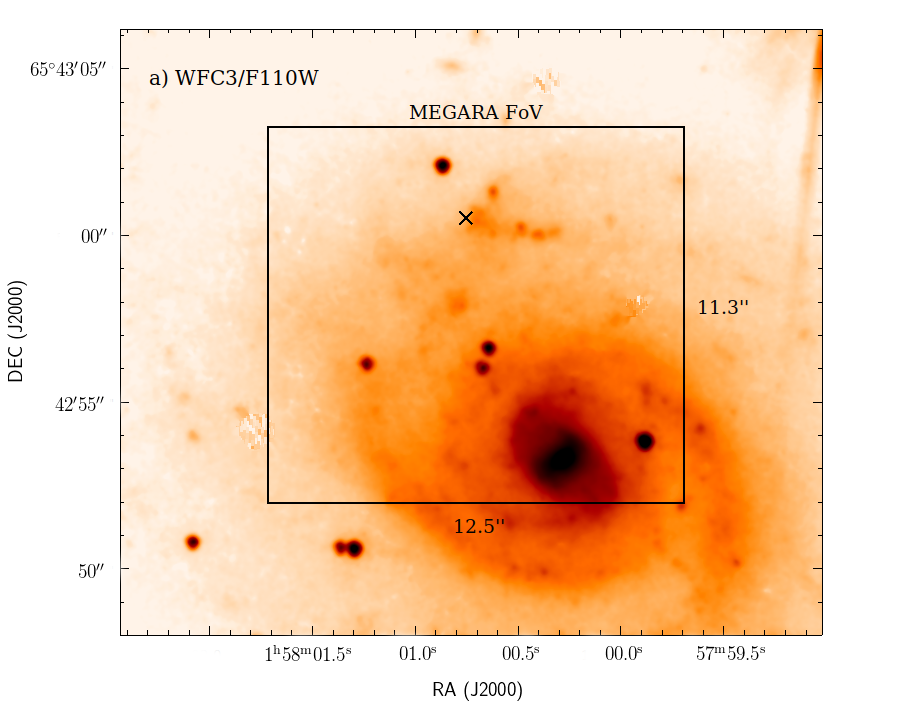}
\includegraphics[width=0.47\linewidth,clip]{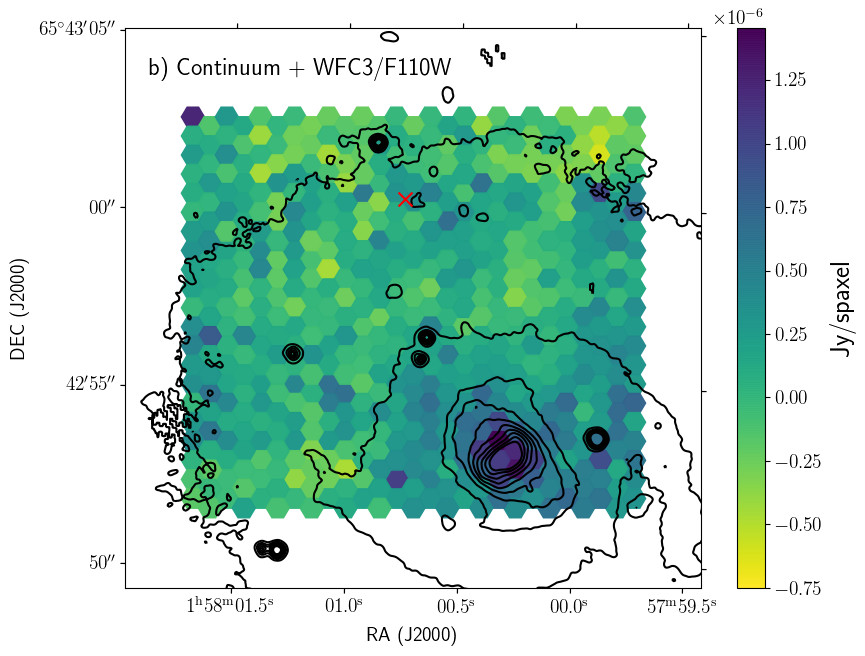}}
\end{minipage}
\begin{minipage}[h]{1.\linewidth}
\center{
\includegraphics[width=0.47\linewidth,clip]{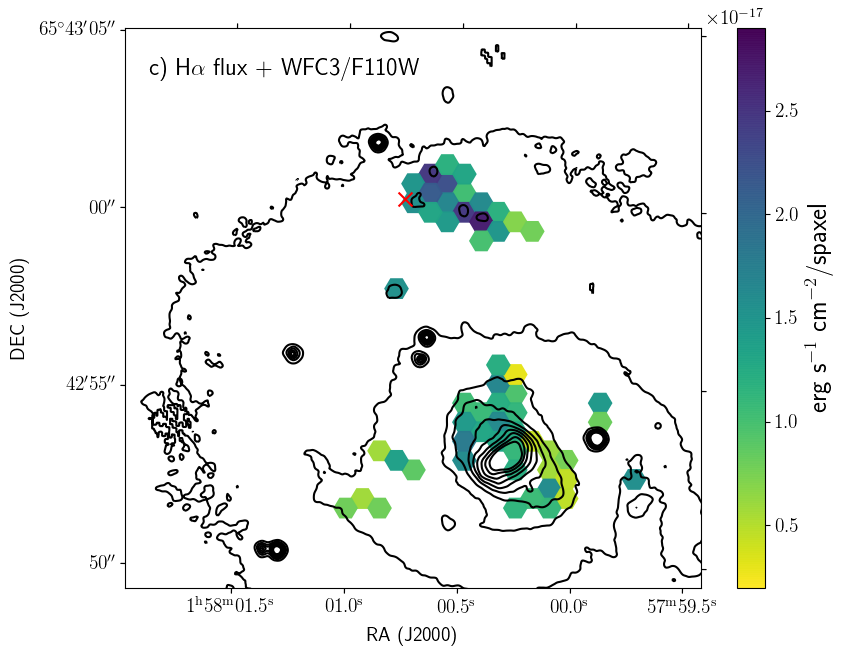}
\includegraphics[width=0.47\linewidth,clip]{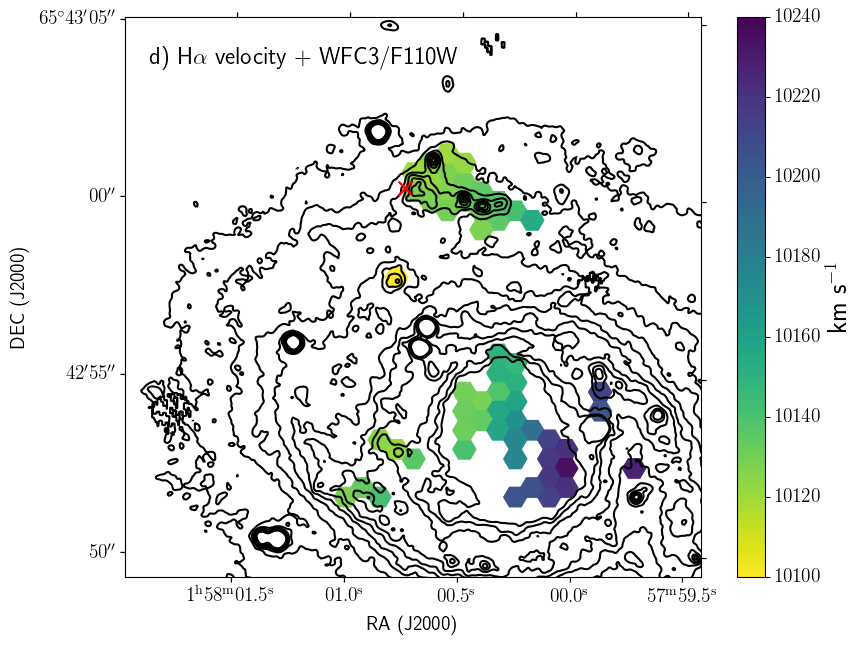}}
\end{minipage}
\begin{minipage}[h]{1.\linewidth}
\center{
\includegraphics[width=0.47\linewidth,clip]{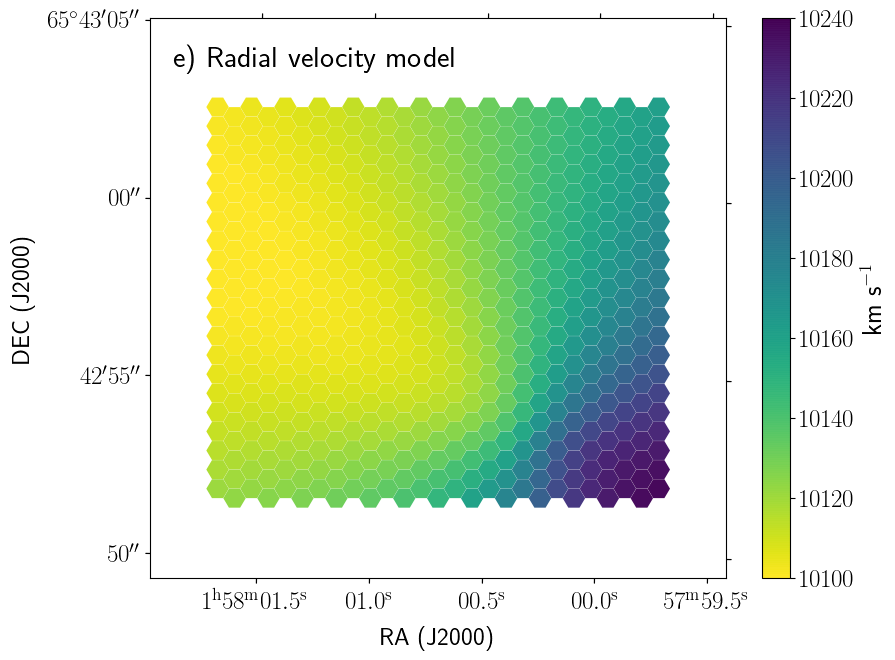}
\includegraphics[width=0.47\linewidth,clip]{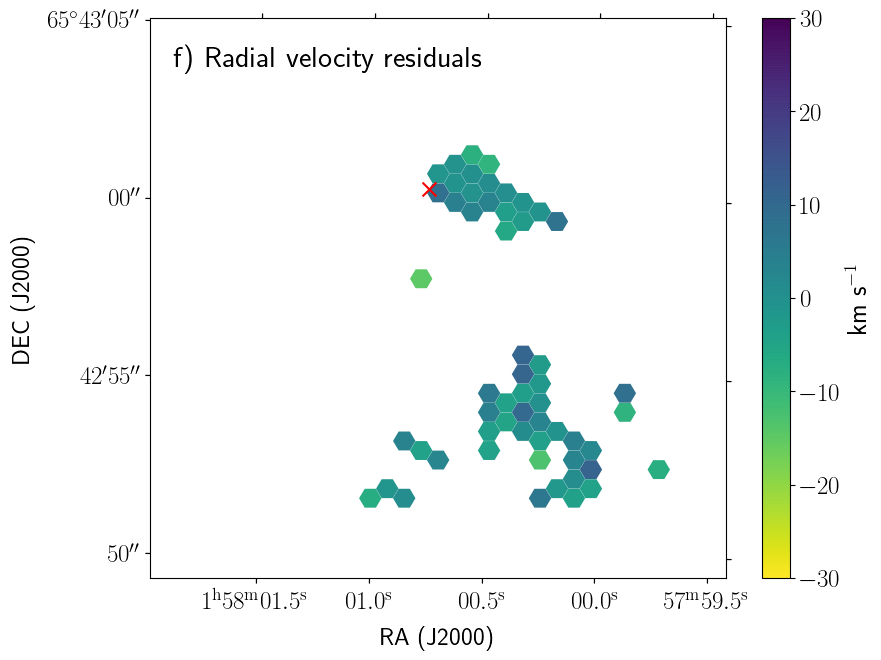}}
\end{minipage}
\caption{{\bf a)} A section of the \hst\ F110W image illustrating the MEGARA FoV. The position of \rthree\ is shown by a cross (as well as on the three subsequent panels). {\bf b)} Average continuum flux level (in Jy per spaxel) from the LR-R setup observations within the spectral range between 6100 and 7200\,\AA. The contours of the F110W image ranging from 0.02 to 0.47\,e$^-$/s/pixel (or between 1.36$\times$10$^{-9}$ and 3.2$\times$10$^{-8}$\,Jy/pixel) in intervals of 0.05\,e$^-$/s/pixel are overlaid; {\bf c)} H$\alpha$ flux of those spaxels with S/N $\geq 5$ at the peak of the line relative to the continuum RMS with the F110W contours overlaid. The contours shown here are identical to those drawn in panel {\bf b}. Note that the three brightest (in H$\alpha$) spaxels in the `V'-shaped region coincide with local maxima in the F110W band. These regions have been used to properly correct the astrometry of the MEGARA LCB data (see text for details); {\bf d)} H$\alpha$ radial velocity data with F110W contours overlaid, ranging from 0.02 to 0.11 in intervals of 0.01\,e$^-$/s. {\bf e)} Best-fitting purely-rotating, inclined thin-disk kinematical model (see text). {\bf f)} Residuals of the radial velocity after the best-fitting thin-disk model has been subtracted from the observed H$\alpha$ radial velocities. The best-fitting velocity RMS is only 5.7\,km\,s$^{-1}$. }
\label{fig:gtc}
\end{figure*}

In panel {\bf b} of Figure~\ref{fig:gtc} we present the distribution of the continuum emission obtained by averaging the flux of the RSS in the wavelength range between 6100 and 7200\,\AA. We note that the fluxes shown here are per spaxel in the case of MEGARA and per pixel in the case of the HST WFC3/F110W image. In order to derive the emission line properties of the area of the \rthreehost\ galaxy covered by our MEGARA LCB observations, we made use of custom Python 3 code based on the \texttt{lmfit} package. This code allows one to simultaneously fit a linear local continuum and the emission line profile (as a Gauss-Hermite series) for each LCB spaxel. The code generates an output RSS where each channel corresponds to a different property (line flux, equivalent width, radial velocity, etc.). In panel {\bf c} we show the results of the analysis of the \halpha\ line adopting a single Gaussian profile for all spaxels with a signal-to-noise ratio at the peak of the line relative to the RMS of the continuum of S/N $\geq 5$. This figure shows, on one hand, the presence of line emission associated with the nuclear spiral and, on the other hand, a series of bright, high-surface-brightness emission-line clumps associated with the brightest regions of the `V'-shaped structure located $\sim$7\arcsec\ north of the galaxy center.  The compactness of these three regions both in HST images and in the MEGARA line-emission data was used to perform a correction of 0\farcs9 east and 1\farcs5 south to the MEGARA astrometry. In panel {\bf d} we provide the radial velocities of the ionized gas as given by \halpha\ for the same S/N $\geq 5$ spaxels. Here we can clearly see that most of the east side of the galaxy shows approaching velocities compared to the galaxy nucleus suggesting that the kinematical minor axis is approximately located in the north-south direction. It is also important to emphasize that the `V'-shaped structure shows a radial velocity in H$\alpha$ that does not differ much from that of the rest of the galaxy, especially if we take into account the fact that the purely rotating gas in that part of the galaxy is moving towards us (see Section~\ref{sec:modelling} for more details). The best-fitting systemic barycentric velocity is $10190^{+8}_{-4}\,\mathrm{km\,s^{-1}}$ leading to the redshift z=0.03399. This result is consistent with that presented by \citet{mnh+20}.

Apart from \halpha\, we detect [NII]$\lambda$6584, [SII]$\lambda$6717 and [SII]$\lambda$6731 lines in the combined spectrum of all spaxels with a signal to noise $>5$ at the peak of the \halpha\ line. The [NII]$\lambda$6584/\halpha\ ratio ($N2 = \log{\mathrm{[NII]/H\alpha}} = -0.745$) can be used to estimate the ionized gas metallicity in this region of the galaxy $12+\log{[O/H]} = 8.4 \equiv \mathrm{Z_\odot}/2$ \citep{marino2013}. In the 19 spaxels to the immediate west of \rthree\ the $N2$ ratio is measured to be $N2=-0.73^{+0.1}_{-0.2}$, consistent with the galaxy-wide average measured above. 

The radial velocity measurement above and the kinematic modeling below focus on \halpha\ since it is brightest and most precisely measured. Measuring the radial velocity differences between the [NII] and \halpha\ lines in individual spaxels, we find that the mean and RMS velocity difference is $1.5\pm8.4\,\mathrm{km\,s^{-1}}$. The RMS is dominated by the radial velocity uncertainties from each [NII] line measurements.


\subsubsection{Kinematic Modelling of the Host Galaxy}
\label{sec:modelling}
\begin{figure}
    \centering
    \includegraphics[width=0.48\textwidth, clip=True, 
    trim={0cm, 0cm, 0cm, 0cm}]{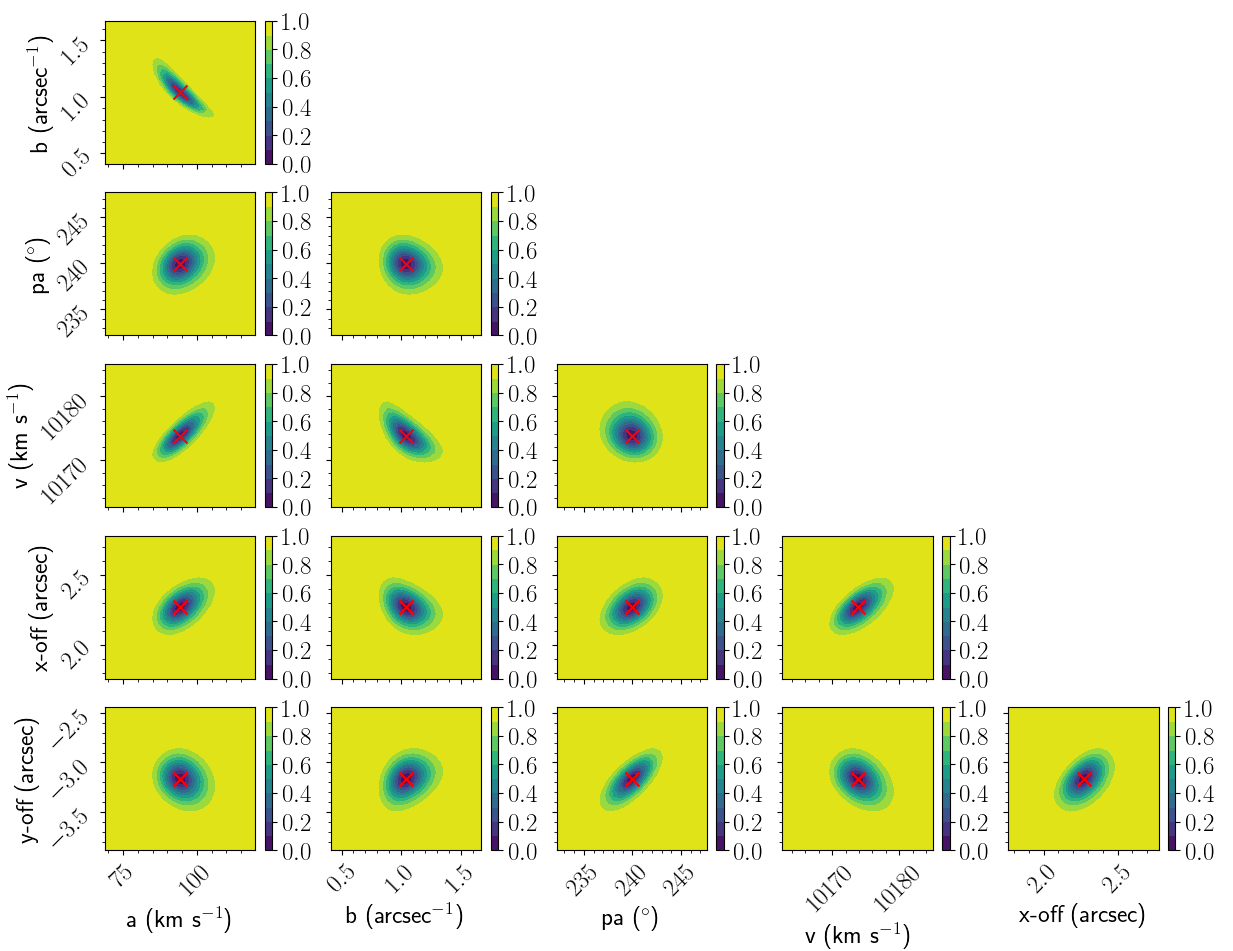}
    \caption{Confidence regions of the disk modeling parameters are shown. The red cross shows the best fit model. The 1$\sigma$ confidence intervals in the text correspond to the 68\% contours in each panel of this corner plot.}
    \label{fig:covariance}
\end{figure}

In order to test the hypothesis that the gas in the `V'-shaped structure located near the FRB position is actually participating in the overall rotation of the gas in the disk, we have built a thin-disk kinematical model assuming a fixed inclination ($i$) at all galactocentric distances $R$  and a rotation curve parameterized as $a \times\,\arctan(b \times R)$. The inclined rotating disk model is then projected onto the sky plane where a radial velocity is calculated for every spaxel's location. We then explored a wide range of parameters (position angle, RA$_\mathrm{center}$, Dec$_\mathrm{center}$, V$_{\mathrm{sys}}$, $a$, $b$) and derived the model that yields the minimum sum of the quadratic differences between data and model. The inclination, which is poorly constrained given the sparse information on radial velocities for this object, was adopted to be the photometric one ($33^\circ\pm6^{\circ}$) as measured from imaging data in the next section after assuming an axisymmetric disk. The rms of the residual obtained is 5.7\,km\,s$^{-1}$ for a kinematical position angle of $239.9^\circ\pm1.5^\circ$ and a best-fitting rotation curve with $a = 94\pm5\,\mathrm{km\,s^{-1}}$ and an inverse “core” radius of $b=1.04^{+0.11}_{-0.13}\,\mathrm{arcsec^{-1}}$ (all errors are 1$\sigma$). Figure~\ref{fig:covariance} shows the covariance in the disk parameters.
The best-fitting model and the corresponding residuals are shown in panels {\bf e} and {\bf f} of Figure~\ref{fig:gtc}, respectively. The homogeneity and low amplitude of the residuals shown in panel {\bf e} indicates that all regions detected in \halpha\ can be reproduced by a simple thin-disk, purely-rotating kinematical model.  

\subsection{High-Resolution Imaging}
\begin{figure*}
    \centering
    \includegraphics[width=\textwidth, clip=True, 
    trim={0cm, 0cm, 0cm, 0cm}]{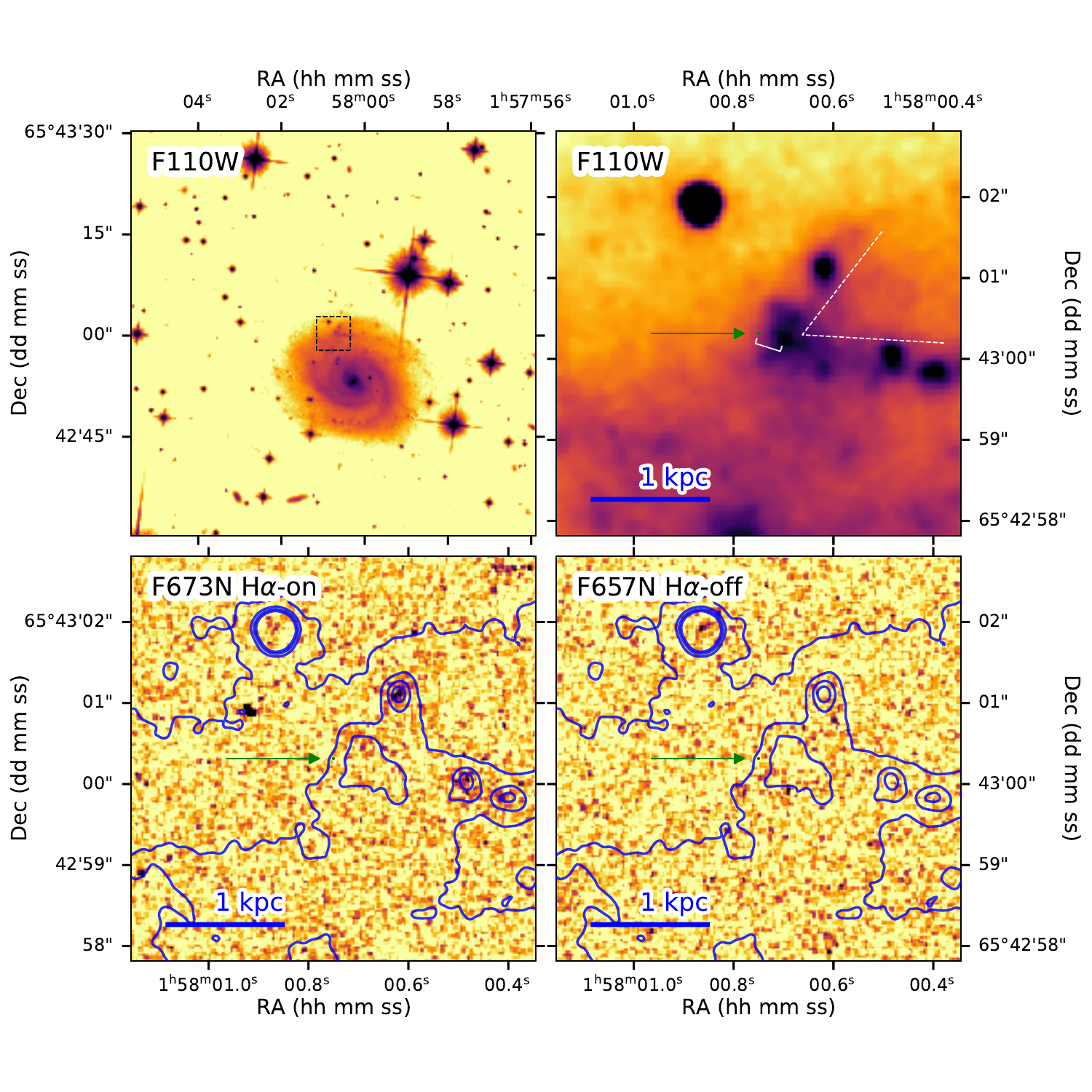}
    \caption{\hst\ observations of \rthree's host galaxy. The 60\arcsec $\times$ 60\arcsec\ F110W image (top, left) shows the full image of \rthreehost\ and its surroundings. The dashed black box denotes the 5\arcsec $\times$ 5\arcsec\ region shown in the zoomed-in images: F110W (top, right), F673N (bottom, left; \halpha-on), and F657N (bottom, right; \halpha-off). The position of the FRB source, including the astrometric uncertainties in its localization and radio-to-optical frame transfer, is shown by the green ellipse at the center of each zoomed-in figure (pointed to by the green arrow). The blue bar indicates the angular scale corresponding to 1\,kpc at the distance of \rthreehost. The F110W zoomed-in image is annotated to show the `V'-shaped structure and the 0\farcs355 separation between \rthree\ and the center of the nearest \halpha\ blob. The F673N and F657N images are overplotted with F110W intensity contours to guide the eye. The color scale of each image is inverted.}
    \label{fig:hst_combined}
\end{figure*}

Figure~\ref{fig:hst_combined} shows the  1\arcmin $\times$ 1\arcmin\ field around \rthree\ in the \hst\ F110W filter (top left) and the 5\arcsec $\times$ 5\arcsec\ zoomed-in fields (marked by the dashed black box) in the F110W (top right), F673N (bottom left, \halpha-on), and F657N (bottom right, \halpha-off) filters. The images are centered at the location of \rthree\ with a green ellipse (pointed to by the green arrow) showing the astrometric uncertainties in the localization and radio-to-optical frame registration of 36\,mas. 

\rthree\ is located off the vertex of the `V'-shaped structure which lies along the spiral arm of \rthreehost. The `V'-shaped structure is indicated in the top left panel of Figure~\ref{fig:hst_combined}. The vertex of the `V' has an  emission region with full width at half maximum (FWHM) size of approximately 0\farcs27, corresponding to about 190\,pc. The region's shape and the underlying background emission has complex structure making it challenging to describe with a single number. This size has not been corrected for the 0\farcs095 resolution of the F110W image --- we expect the region to be more compact. The region is not bright in \halpha\ and is barely detectable in the F673N image. The \halpha\ luminosity and the star formation surface density of the vertex is discussed in Section~\ref{sec:star-formation}. We assume that the F110W light also traces the \halpha\ distribution within the vertex region, hence the brightest F110W pixel likely has the highest star-formation density.

\rthree\ is 0\farcs355, i.e. $\sim 250$\,pc away from the brightest pixel in the F110W image. The offset is similar to the $\sim 200$\,pc separation of \rone\ from the center of the \halpha\ knot in its host galaxy \citep{bassa2017frb, kokubo2017}. The 380-pc size of the star-forming region for \rthree\ is much smaller than the 1.4--1.9\,kpc size of the star-forming knot hosting \rone\ \citep{bassa2017frb, kokubo2017} . The magnitude of this offset compared to the offsets of other compact objects (either isolated or in binaries) from their birth places sets strong constraints on the age and nature of \rthree, as discussed in Section~\ref{sec:discussion}.

\rthree\ is located in the Milky Way plane towards the anti-center. The Bayestar19 \citep{green2019} estimate for the reddening between the $g^\prime$ and $r^\prime$ filters is $E_{g-r} = 0.69\pm0.02$ (1-$\sigma$) based on the PanSTARRs, 2MASS, and \emph{Gaia} data. We follow the scaling prescribed by \citet{green2019}, and the recalibrated extinction law in \citet{schlafly2011}, to estimate the extinction in the F110W and the F673N filters to be 0.61\,mag and 1.37\,mag, respectively\footnote{We note that the older extinction map from \citet{schlafly2011}, based on SDSS photometry, estimates the F110W extinction to be 0.87\,mag along this line-of-sight. The discrepancy between the older and newer estimates does not qualitatively affect our conclusions, and hence we use the newer estimates from \citet{green2019}.}. 

The host galaxy extinction is assumed to be negligible, considering its nearly face-on orientation, for the IR wide-band imaging (F110W). However, the conversion of \halpha\ flux to star-formation rate includes a correction for the typical host extinction \citep[][]{kewley2002}. 

We also performed a surface photometry analysis on the F110W HST image using the \texttt{photutils.isophote} package of \texttt{Astropy}. The package estimates the isophotes using the method described in \cite{1987MNRAS.226..747J}. By assuming an intrinsically axisymmetric, infinitesimally thin disk, we estimate the inclination angle of the stellar disk, $i_{\rm{stellar}} = 33^{\circ} \pm 6^{\circ}$ (1-$\sigma$), which is in agreement with the inclination angle estimated using the \halpha\ data. In estimating $i_{\rm{stellar}}$, we only considered the projected galactocentric radius range between 4\,kpc and 10\,kpc because at radii $< 4$\,kpc, the radial profile traces the bulge of the galaxy and beyond 10\,kpc the fitted ellipticity values show large swings.

\subsection{Star Formation}
\label{sec:star-formation}
The 5-$\sigma$ limit for \halpha\ emission from a point source at the \rthree\ location (above the diffuse emission of the host galaxy)  is 25.42\,mag (Vega). Assuming the continuum to be negligible and all the light to be due to the redshifted-\halpha\ photons, this corresponds to a flux limit of $3\times10^{-18}\,\mathrm{erg\,cm^{-2}\,s^{-1}}$ after correcting for the Milky Way extinction. This corresponds to a point-source \halpha\ luminosity of $8\times10^{36}\,\mathrm{erg\,s^{-1}}$. Using the conversions of \citet{kewley2002}, the star formation rate at the location of \rthree\ is constrained to be $\lesssim10^{-4}\,\mathrm{\msun\,yr^{-1}}$. The resolution ($\lambda/D$) of the F673N image is $\approx58\,$mas. Given this size scale, the star formation surface density at the location of \rthree\ should be $\lesssim2\times10^{-2}\,\mathrm{\msun\,yr^{-1}\,kpc^{-2}}$.

The nebular region at the vertex of the `V' is measurably extended in the F110W and F673N images. To measure the total star formation rate in the region, we  smoothed the F673N image with a Gaussian kernel with a $\sigma=125$\,mas revealing a detectable blob with a total extinction-corrected flux of $\approx3\times10^{-17}\,\mathrm{erg\,cm^{-2}\,s^{-1}}$, corresponding to an \halpha\ luminosity $\approx9\times10^{37}\,\mathrm{erg\,s^{-1}}$ and a star formation surface density of $3\times10^{-2}\,\mathrm{\msun\,yr^{-1}\,kpc^{-2}}$. The measurement assumes a photometric zeropoint at infinite radius. This is roughly consistent with the extinction-corrected \halpha\ flux of $5.5\times10^{-17}\,\mathrm{erg\,s^{-1}}$ measured in the 620-mas-wide MEGARA spaxel located at the vertex of the `V'. The difference in the flux measurement is likely due to extra emission outside the \hst\ photometric region in the wings of the \halpha\ clump.

\subsection{\ion{H}{2} Regions and  O/B stars}
In the \hst\ F110W image, we constrain a point-source at the location of \rthree\ to be fainter than 27.15\,mag (Vega). Including the extinction correction discussed above and the distance modulus of \rthreehost, this corresponds to an absolute magnitude limit of $M_{F110W} > -7.97$\,mag (Vega). The J-band magnitude of a O3V star is $\approx-4.9$\,mag \citep{worthey2011, pecaut2013}. Thus, the F110W data are unable to constrain the presence of single bright stars. 

However, the upper limit on the \halpha\ luminosity at the location of \rthree\ can constrain the rate of ionizing photons in the neighborhood. \cite{gutierrez2011properties} demonstrated a tight correlation between the \halpha\ luminosity $L_\alpha$ and radius for \ion{H}{2} regions in M51 and NGC 4449. An \ion{H}{2} region with $L_\alpha < 10^{37}\,\mathrm{erg\,s^{-1}}$ is expected to be smaller than 10--60\,pc (including the scatter in the relation). Consequently, we do not expect that our non-detection of an \ion{H}{2} region in the F673N image is because the \halpha\ flux is resolved out. 

The $L_\alpha$ limit can be converted to a limit on the hydrogen ionizing flux $Q(H^0)<9\times10^{48}\,\mathrm{s^{-1}}$ \citep[see][for conversion constants]{osterbrock2006astrophysics}. From the $Q(H^0)$ calculations of \citet{martins2005}, we can rule out a single main sequence star hotter than O6V. For giants and supergiants, stars hotter than O7.5III and all OI stars are ruled out. 

\section{Discussion}
\label{sec:discussion}
Our observations and constraints on the environment of \rthree\ set it apart from the other localized FRBs, and challenge some of the theoretical models put forward to explain its periodic activity. Here we discuss the observational and theoretical implications of these constraints.

\subsection{Comparison to FRB\,20190608B}
FRB\,20190608B \citep[previously FRB 190608;][]{macquart2020} is an apparently non-repeating FRB that was detected and localized by ASKAP to a spiral host galaxy, SDSS\,J221604.90$-$075356.0, at a redshift of $z=0.11778$. The location of FRB\,20190608B in the spiral arm of SDSS\,J221604.90$-$075356.0, a face-on spiral, is strikingly similar to that of \rthree. \citet{chittidi2020} acquired integral field spectra and \hst\ imaging of the host galaxy and measured a local star formation surface density of $1.2\times10^{-2}\,\mathrm{\msun\,yr^{-1}\,kpc^{-2}}$ at the location of FRB\,20190608B. This is similar to the star formation density in the \halpha\ blob at the vertex of the `V'-shaped structure but significantly higher than the star formation density at the location of \rthree. The localization precision of FRB\,20190608B of $\approx0\farcs26$ (1-$\sigma$) corresponds to a physical scale of 0.55\,kpc at the redshift of $z=0.11778$. Hence, any offset from the star-forming region, similar to that seen for \rthree\ and \rone, cannot be measured unless repeat bursts are detected and localized with milliarcsecond precision. Similar to FRB\,20190608B, \rthree\ is also not found to be coincident with or near the brightest star-forming region in the host galaxy. While \citet{chittidi2020} noted the spectral and the mass differences between the host galaxies, the local environments of \rthree, a repeater, and FRB\,20190608B, an as-yet non-repeater\footnote{\citet{day2020} have shown that FRB\,20190608B showed some properties similar to repeater bursts --- it had a high degree of linear polarization, a non-varying position angle through the burst, and possible complex structure underneath the scattering tail.}, are very similar. Continued monitoring for repeat bursts from FRB\,20190608B would help to improve its localization and to refine the comparison of its nature with that of \rthree. 


\subsection{Nature of \rthree}
The high-resolution, multi-band optical imaging and spectroscopy we present here provide important insights into the nature of \rthree, which complement what can be discerned from the spectro-temporal and polarimetric properties of the bursts themselves --- as well as the periodic activity of the source. We first summarize what was known previously, and then discuss the implications of the new results we present here.

\subsubsection{Previous results}

Observations of $\sim$3--4\,{\textmu}s burst structure place tight constraints on the size of the emitting region \citep{nimmo2020}; in the absence of special relativistic effects, this corresponds to a $\sim$1\,km region, given light-crossing-time arguments.  In the context of magnetar models, this short timescale, along with the range of observed temporal timescales spanning 3--4 orders of magnitude from $\sim${\textmu}s$-$ms \citep{nimmo2020}, is more naturally explained in terms of emission generated relatively close to the neutron star \citep{beniamini2020b, lyutikov2020, lu2020} --- as opposed to much further out in a relativistic shock \citep{mm18, bel17}.

The 16.35-day activity period is also a key insight, and differentiates \rthree\ from the isolated Galactic magnetar, and putative FRB source, \sgr. If FRB 20180916B is produced by a strongly magnetised neutron star, some extra ingredient is necessary to understand the emission mechanism. The activity period could in principle be related to rotation \citep{beniamini2020a}, precession \citep{levin2020,zanazzi2020,sobyanin2020,yang2020} or an orbit \citep{mvz20,lyutikov2020,iokazhang2020}. The near constancy of polarization angle within and {\it between} bursts places strong constraints on precession models \citep{nimmo2020}.  The similar constraints imposed by the constant polarization angle of \rone\ \citep{msh+18} argue that precession models are disfavored. The variation in rotation measure, which may correlate with orbital phase \citep{pleunis2020}, suggests the presence of a variable magneto-ionic medium around the system, which is naturally explained in an orbiting binary model.  See \citet{pleunis2020} also for a longer discussion of how the observed frequency dependence of observed burst activity could be interpreted in the context of a binary model.

\subsubsection{Constraints from this work}
First, the radial velocity measurements and kinematic modeling from the MEGARA observations show that the `V'-shaped structure is dynamically a part of the spiral galaxy and excludes the possibility that the ionized gas that we detect belongs to a  satellite galaxy --- a possibility discussed in \citet{mnh+20}, when only seeing-limited images and single-slit spectroscopy of the galaxy were available.

Our \halpha\ on/off observations constrain the \halpha\ luminosity of an unresolved \ion{H}{2} region at the location of \rthree\ to be $<8\times10^{36}\,\mathrm{erg\,s^{-1}}$. The \halpha\ luminosities of \ion{H}{2} regions range from $10^{34-38}\,\mathrm{erg\,s^{-1}}$ \citep{fich1990, azimlu2011}, with the `knee' of the distribution being $10^{36.7}\,\mathrm{erg\,s^{-1}}$. Thus we can rule out the brightest \ion{H}{2} regions powered by the youngest massive stars. Specifically, based on the rate of ionizing photons and the corresponding \halpha\ luminosity, we can constrain a possible stellar companion to \rthree\ to be cooler and smaller than O6V and O7.5III spectral types. All supergiant O stars can be ruled out. \halpha\ emission line stars (late O or B spectral types), which have typical \halpha\ luminosities of $10^{32}$--$10^{34}\,\mathrm{erg\,s^{-1}}$ \citep{apparao1997}, cannot be ruled out by these observations.

\begin{figure}
    \centering
    \includegraphics[width=\columnwidth, clip=True, 
    trim={0cm, 0cm, 0cm, 0cm}]{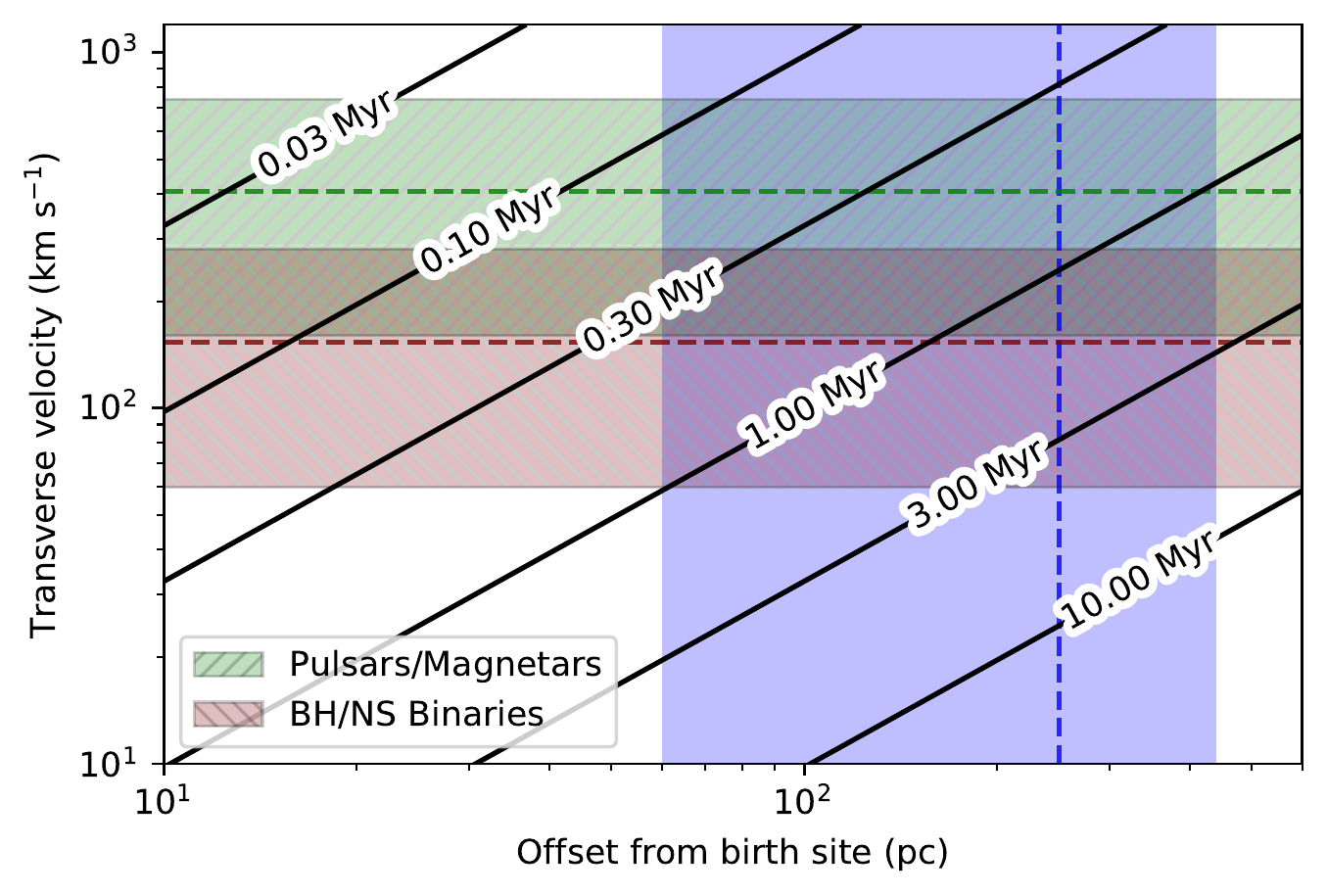}
    \caption{Constraints on the age of \rthree\ based on the proper motions of isolated neutron stars and those in binaries. The transverse offset from the peak of the nearest \halpha\ blob (vertical dashed blue line) and the range of offsets from the presumed birth site (blue region) are shown. The range is determined by the full width at half maximum of the blob size in the F110W image along the line joining the FRB location to the brightest pixel. The 90\% range of 1-D velocities of pulsars \citep{hobbs2005}, magnetars \citep{tendulkar2014, ding2020} and neutron stars in binaries \citep{bodaghee2012}, assuming a Maxwellian distribution, are also shown (green and maroon regions, respectively). The median velocities are shown by dashed lines. Diagonal black lines indicate the ages corresponding to a given offset and velocity.}
    \label{fig:progenitor_age}
\end{figure}

Our HST observations demonstrate that \rthree\ is offset by $250 \pm 190$\,pc from the nearest knot of active star formation in the host galaxy. The separation is measured from the brightest pixel of the F110W image. \citet{bassa2017frb} found a comparably large, $\sim 200$\,pc offset for \rone. This is problematic for models that require a young magnetar (age of $\sim$10--$10^5$\,yr), since such sources are invariably found close to their birth sites. For comparison, the scale height of Galactic magnetars is only 20--30\,pc \citep{ok14}. While scale heights are measured perpendicular to the plane of the Galaxy and the offset measured for \rthree\ is in the plane of its host, since natal kicks for magnetars are statistically isotropic, the comparison is valid. Figure~\ref{fig:progenitor_age} shows the expected age of \rthree\ given the range of possible offsets and the typical velocities of pulsars, magnetars, and X-ray binaries in the Milky Way. With the typical projected velocities of pulsars, magnetars, or neutron stars in binaries ($60-750$\,km~s$^{-1}$), \rthreefull\ would need 800\,kyr to 7\,Myr to traverse the observed distance from its presumed birth site. Even with a kick velocity of $\sim1000$\,km~s$^{-1}$ at birth, a neutron star would still require $\sim0.25$\,Myr to traverse the 250\,pc offset we determine for \rthree. 

It is possible that a putative young magnetar was born at the edge of the star-forming region and did not have to travel very far from its birth site. However, such a magnetar is more likely to originate in a region with a higher density of massive star progenitors (i.e. the center of the cluster) rather than at the edges. It is not straightforward to quantify this probability distribution based on the F110W and F673N images. Hence, we show only the distance from the brightest pixel and an approximate size for the cluster in Figure~\ref{fig:progenitor_age}, but the probability distribution is not uniform across the range.

Though magnetars may also be created in compact-binary mergers or accretion-induced collapse --- as opposed to the core-collapse of a massive star --- such channels have a much lower rate, and thus our findings suggest that \rthree\ is associated with a much older, $\sim$Myr, neutron star. Nonetheless, the relative proximity of star-formation --- in fact a very striking complex of star-forming regions in the host galaxy --- is unlikely to be pure coincidence, and suggests that \rthree\ is not a very old ($\gtrsim10-50$\,Myr) system.  

Another possibility is that a neutron star/magnetar is created {\it in situ} from the supernova of a runaway OB star \citep[see ][ for a review]{zinnecker2007} --- a massive star ejected from a dense stellar cluster through binary interactions at high velocities ($>30\,\mathrm{km\,s^{-1}}$). The neutron star/magnetar would be observed to be well-separated from the stellar cluster but be young enough to create its own energetic phenomena. However, the challenges with this channel are two-fold: First, runaway OB stars represent a small fraction of the population of OB stars \citep[1--10\%; ][]{fujii2011}, and magnetars are a small fraction of the population of neutron stars \citep[10\% of core-collapse rate; see ][]{kaspi2017magnetars}, so this channel has low likelihood. Secondly, given the speeds of these runaway stars (30--120\,km\,s$^{-1}$), they would still require many millions of years to achieve the observed physical offset which is a significant fraction of, if not greater than, the lifetimes of O stars. However, B stars which have longer lifetimes are still possible.  

The Galactic population of high-mass X-ray binaries \citep[HMXBs;][]{walter2015} and $\gamma$-ray binaries \citep{dubus2013} present a potentially interesting analogy.  These systems feature a neutron star and high-mass O- or Be-star companion and have orbital periods that are comparable to \rthree's 16.35-day activity period --- e.g., 1FGL~J1018.6$-$5856 ($P_{\rm orb} = 16.6$\,day) and LS I +61$^{\circ}$303 ($P_{\rm orb} = 26.5$\,day).  \citet{bodaghee2012} consider the spatial correlation of HMXBs and active OB star-forming complexes in the Milky Way.   They find that the locations of HMXBs reflect the distributions of the massive star-forming regions that are expected to produce them.  However, they also determine  an average offset of $0.4\pm0.2$\,kpc between HMXBs and OB associations, which they attribute to natal kick velocities of $100\pm50$\,km~s$^{-1}$ and typical system ages of $\sim4$\,Myr.

\citet{safarzadeh2020} analysed the star formation rate and the separations of FRBs from the centers of their host galaxies (host offsets) for 10 FRBs with secure host associations and compared them to a simple model where FRB rates are proportional to the recent star formation rate (as expected for prompt magnetars). They reported that the star formation rates measured from the host galaxies are incompatible with such a model but the host offset distribution is compatible. \citet{bochenek2020} did a similar study comparing the FRB hosts properties and offsets to those of core-collapse supernovae and showed that the star-formation rates and stellar masses of localized FRB hosts as well as host offsets can be consistent with an origin in magnetars formed from core-collapse supernovae. However, we note that the population of young neutron star binaries would follow the star formation rate and stellar mass distributions to a similar degree as magnetars born in core-collapse supernovae. This highlights the need for precision localization of FRBs: it is not sufficient to know that FRBs occur near star formation sites; we need to understand exactly {\it how near} they are located. 

Interestingly, a 300-ms duration magnetar-like X-ray burst was detected in the direction of the $\gamma$-ray binary LSI +61$^\circ$303 \citep{torres2012} (though there is a non-zero probability that the X-ray burst came from an unrelated background magnetar). While the nature of the compact object in LSI +61$^\circ$303 is still debated, the detection of 269.196-ms radio pulsations from this direction suggests the presence of a neutron star in the binary \citep{lsi61303_atel}. This suggests that the magnetar-vs-binary scenarios are not exclusive. Indeed, the case has been made for magnetars in high-mass X-ray binaries \citep{popov2016}, superfast X-ray transients \citep[SFXTs;][]{bozzo2008}, and ultra-luminous X-ray sources \citep[ULXs;][]{tsygankov2016}. 

Thus, the activity period, positional offset and constraints on local emission are fully consistent with a picture in which \rthree\ is a neutron-star HMXB or $\gamma$-ray binary system with a late O-type or B-type companion.  In such a scenario, interaction between the neutron star's magnetic field and the ionised wind of the companion star may be key to creating the observed radio bursts (FRBs) themselves.  Such interactions could create magnetic re-connections that provide the necessary ingredients to produce coherent radio bursts on a wide range of (apparent) timescales and luminosities.  The observed periodic activity could then be a reflection of the observer's line-of-sight.   The companion wind will compress the magnetosphere of the neutron star on the companion-facing side and create a tail on the opposite side.  An observer may then only see radio bursts when this magnetic tail is pointed towards Earth \citep{iz20}.  Alternatively, such systems are often found to have very high eccentricity ($e = $0.1--0.9), and the variable distance between the neutron star and companion could mean that the companion's wind only strongly compresses the neutron star's magnetosphere at certain orbital phases.

Other interacting binary models have also been proposed, including ones in which a stellar companion wind is generated by a millisecond pulsar \citep{iz20}, or in which the neutron star's magnetosphere is interacting with orbiting asteroids \citep{mvz20}.  We cannot exclude such scenarios, since such systems can also satisfy the observed spatial offset.  However, such systems could potentially be quite old ($\gtrsim10-100$\,Myr), and thus not naturally explain the relative proximity of \rthree\ to such a prominent complex of star formation in the host galaxy.

High-cadence searches for bright radio bursts from Galactic HMXBs and gamma-ray binaries can help to better establish a connection to \rthree.  There are $\sim90$ confirmed and suspected Be/X-ray binaries, $\gamma$-ray binaries and $\approx131$ known HMXBs in the Milky Way and Magellanic Clouds \citep{reig2011}, at distances of up to 50\,kpc.  At such distances, these sources would produce few hundred MJy-ms bursts, if the luminosity is comparable to the weakest bursts seen from \rthree.  Small ($\lesssim25$\,m) radio dishes and individual radio antennas are more than sensitive enough to detect such emission after surmounting the challenge of distinguishing bright astrophysical bursts from radio frequency interference \citep{tkp16}. 

\acknowledgments

J.W.T.H. acknowledges funding from an NWO Vici grant (``AstroFlash''; VI.C.192.045). A.G.P. acknowledges financial support from the Spanish Ministry of Economy and Competitiveness (MINECO) under grant RTI2018-096188-B-I00, which is partly funded by the European Regional Development Fund (ERDF).  M.B. and P.C. are supported by FRQNT Doctoral Research Awards. V.M.K. holds the Lorne Trottier Chair in Astrophysics \& Cosmology and a Distinguished James McGill Professorship and receives support from an NSERC Discovery Grant and Herzberg Award, from an R. Howard Webster Foundation Fellowship from the CIFAR, and from the FRQNT Centre de Recherche en Astrophysique du Qu\'{e}bec. P.S. is a Dunlap Fellow and an NSERC Postdoctoral Fellow. The Dunlap Institute is funded through an endowment established by the David Dunlap family and the University of Toronto. B.M. acknowledges the support by the Spanish Ministerio de Ciencia e Innovaci\'{o}n (MICINN) under grant PID2019-105510GB-C31 and through the ``Center of Excellence Mar\'{i}a de Maeztu 2020-2023'' award to the ICCUB (CEX2019-000918-M). F.K. acknowledges support by the Swedish Research Council. FRB research at UBC is supported by an NSERC Discovery Grant and by the Canadian Institute for Advanced Research. The NANOGrav project receives support from National Science Foundation (NSF) Physics Frontiers Center award number 1430284.

This research is based on observations made with the NASA/ESA Hubble Space Telescope obtained from the Space Telescope Science Institute, which is operated by the Association of Universities for Research in Astronomy, Inc., under NASA contract NAS 5–26555. These observations are associated with program 16072. This research is also based on observations made with the Gran Telescopio Canarias (GTC), installed at the Spanish Observatorio del Roque de los Muchachos of the Instituto de Astrof\'{i}sica de Canarias, in the island of La Palma. This work is partly based on data obtained with MEGARA instrument, funded by European Regional Development Funds (ERDF), through Programa Operativo Canarias FEDER 2014-2020. This work has made use of data from the European Space Agency (ESA) mission {\it Gaia} (\url{https://www.cosmos.esa.int/gaia}), processed by the {\it Gaia} Data Processing and Analysis Consortium (DPAC, \url{https://www.cosmos.esa.int/web/gaia/dpac/consortium}). Funding for the DPAC has been provided by national institutions, in particular the institutions participating in the {\it Gaia} Multilateral Agreement. Part of this research was carried out at the Jet Propulsion Laboratory, California Institute of Technology, under a contract with the National Aeronautics and Space Administration.

%

\vspace{5mm}
\facilities{HST (WFC3), GTC (MEGARA)}

\software{Astropy \citep{astropy:2013, astropy:2018}, DOLPHOT \citep{dolphot2016}, scipy \citep{2020SciPy-NMeth}, numpy \citep{2020NumPy-Array}, LMFIT \citep{lmfit}} 




\bibliography{r3_hst, frbrefs}

\begin{thebibliography}{}
\expandafter\ifx\csname natexlab\endcsname\relax\def\natexlab#1{#1}\fi
\providecommand{\url}[1]{\href{#1}{#1}}
\providecommand{\dodoi}[1]{doi:~\href{http://doi.org/#1}{\nolinkurl{#1}}}
\providecommand{\doeprint}[1]{\href{http://ascl.net/#1}{\nolinkurl{http://ascl.net/#1}}}
\providecommand{\doarXiv}[1]{\href{https://arxiv.org/abs/#1}{\nolinkurl{https://arxiv.org/abs/#1}}}

\bibitem[{{Apparao} \& {Tarafdar}(1997)}]{apparao1997}
{Apparao}, K. M.~V., \& {Tarafdar}, S.~P. 1997, Bulletin of the Astronomical
  Society of India, 25, 345

\bibitem[{{Astropy Collaboration} {et~al.}(2013){Astropy Collaboration},
  {Robitaille}, {Tollerud}, {Greenfield}, {Droettboom}, {Bray}, {Aldcroft},
  {Davis}, {Ginsburg}, {Price-Whelan}, {Kerzendorf}, {Conley}, {Crighton},
  {Barbary}, {Muna}, {Ferguson}, {Grollier}, {Parikh}, {Nair}, {Unther},
  {Deil}, {Woillez}, {Conseil}, {Kramer}, {Turner}, {Singer}, {Fox}, {Weaver},
  {Zabalza}, {Edwards}, {Azalee Bostroem}, {Burke}, {Casey}, {Crawford},
  {Dencheva}, {Ely}, {Jenness}, {Labrie}, {Lim}, {Pierfederici}, {Pontzen},
  {Ptak}, {Refsdal}, {Servillat}, \& {Streicher}}]{astropy:2013}
{Astropy Collaboration}, {Robitaille}, T.~P., {Tollerud}, E.~J., {et~al.} 2013,
  \aap, 558, A33, \dodoi{10.1051/0004-6361/201322068}

\bibitem[{{Astropy Collaboration} {et~al.}(2018){Astropy Collaboration},
  {Price-Whelan}, {Sip{H{o}}cz}, {G{"u}nther}, {Lim}, {Crawford}, {Conseil},
  {Shupe}, {Craig}, {Dencheva}, {Ginsburg}, {Vand erPlas}, {Bradley},
  {P{'e}rez-Su{'a}rez}, {de Val-Borro}, {Aldcroft}, {Cruz}, {Robitaille},
  {Tollerud}, {Ardelean}, {Babej}, {Bach}, {Bachetti}, {Bakanov}, {Bamford},
  {Barentsen}, {Barmby}, {Baumbach}, {Berry}, {Biscani}, {Boquien}, {Bostroem},
  {Bouma}, {Brammer}, {Bray}, {Breytenbach}, {Buddelmeijer}, {Burke},
  {Calderone}, {Cano Rodr{'i}guez}, {Cara}, {Cardoso}, {Cheedella}, {Copin},
  {Corrales}, {Crichton}, {D'Avella}, {Deil}, {Depagne}, {Dietrich}, {Donath},
  {Droettboom}, {Earl}, {Erben}, {Fabbro}, {Ferreira}, {Finethy}, {Fox},
  {Garrison}, {Gibbons}, {Goldstein}, {Gommers}, {Greco}, {Greenfield},
  {Groener}, {Grollier}, {Hagen}, {Hirst}, {Homeier}, {Horton}, {Hosseinzadeh},
  {Hu}, {Hunkeler}, {Ivezi{'c}}, {Jain}, {Jenness}, {Kanarek}, {Kendrew},
  {Kern}, {Kerzendorf}, {Khvalko}, {King}, {Kirkby}, {Kulkarni}, {Kumar},
  {Lee}, {Lenz}, {Littlefair}, {Ma}, {Macleod}, {Mastropietro}, {McCully},
  {Montagnac}, {Morris}, {Mueller}, {Mumford}, {Muna}, {Murphy}, {Nelson},
  {Nguyen}, {Ninan}, {N{"o}the}, {Ogaz}, {Oh}, {Parejko}, {Parley}, {Pascual},
  {Patil}, {Patil}, {Plunkett}, {Prochaska}, {Rastogi}, {Reddy Janga},
  {Sabater}, {Sakurikar}, {Seifert}, {Sherbert}, {Sherwood-Taylor}, {Shih},
  {Sick}, {Silbiger}, {Singanamalla}, {Singer}, {Sladen}, {Sooley},
  {Sornarajah}, {Streicher}, {Teuben}, {Thomas}, {Tremblay}, {Turner},
  {Terr{'o}n}, {van Kerkwijk}, {de la Vega}, {Watkins}, {Weaver}, {Whitmore},
  {Woillez}, {Zabalza}, \& {Astropy Contributors}}]{astropy:2018}
{Astropy Collaboration}, {Price-Whelan}, A.~M., {Sip{H{o}}cz}, B.~M., {et~al.}
  2018, aj, 156, 123, \dodoi{10.3847/1538-3881/aabc4f}

\bibitem[{{Azimlu} {et~al.}(2011){Azimlu}, {Marciniak}, \&
  {Barmby}}]{azimlu2011}
{Azimlu}, M., {Marciniak}, R., \& {Barmby}, P. 2011, \aj, 142, 139,
  \dodoi{10.1088/0004-6256/142/4/139}

\bibitem[{{Bannister} {et~al.}(2019){Bannister}, {Deller}, {Phillips},
  {Macquart}, {Prochaska}, {Tejos}, {Ryder}, {Sadler}, {Shannon}, {Simha},
  {Day}, {McQuinn}, {North-Hickey}, {Bhandari}, {Arcus}, {Bennert}, {Burchett},
  {Bouwhuis}, {Dodson}, {Ekers}, {Farah}, {Flynn}, {James}, {Kerr}, {Lenc},
  {Mahony}, {O{\textquoteright}Meara}, {Os{\l}owski}, {Qiu}, {Treu}, {U},
  {Bateman}, {Bock}, {Bolton}, {Brown}, {Bunton}, {Chippendale}, {Cooray},
  {Cornwell}, {Gupta}, {Hayman}, {Kesteven}, {Koribalski}, {MacLeod},
  {McClure-Griffiths}, {Neuhold}, {Norris}, {Pilawa}, {Qiao}, {Reynolds},
  {Roxby}, {Shimwell}, {Voronkov}, \& {Wilson}}]{bdp+19}
{Bannister}, K.~W., {Deller}, A.~T., {Phillips}, C., {et~al.} 2019, Science,
  365, 565, \dodoi{10.1126/science.aaw5903}

\bibitem[{Bassa {et~al.}(2017)Bassa, Tendulkar, Adams, Maddox, Bogdanov, Bower,
  Burke-Spolaor, Butler, Chatterjee, Cordes, {et~al.}}]{bassa2017frb}
Bassa, C., Tendulkar, S., Adams, E., {et~al.} 2017, The Astrophysical Journal
  Letters, 843, L8

\bibitem[{{Beloborodov}(2017)}]{bel17}
{Beloborodov}, A.~M. 2017, \apjl, 843, L26, \dodoi{10.3847/2041-8213/aa78f3}

\bibitem[{{Beniamini} \& {Kumar}(2020)}]{beniamini2020b}
{Beniamini}, P., \& {Kumar}, P. 2020, \mnras, 498, 651,
  \dodoi{10.1093/mnras/staa2489}

\bibitem[{{Beniamini} {et~al.}(2020{\natexlab{a}}){Beniamini}, {Wadiasingh}, \&
  {Metzger}}]{beniamini2020}
{Beniamini}, P., {Wadiasingh}, Z., \& {Metzger}, B.~D. 2020{\natexlab{a}},
  \mnras, 496, 3390, \dodoi{10.1093/mnras/staa1783}

\bibitem[{{Beniamini} {et~al.}(2020{\natexlab{b}}){Beniamini}, {Wadiasingh}, \&
  {Metzger}}]{beniamini2020a}
---. 2020{\natexlab{b}}, \mnras, 496, 3390, \dodoi{10.1093/mnras/staa1783}

\bibitem[{{Bhandari} {et~al.}(2020{\natexlab{a}}){Bhandari}, {Bannister},
  {Lenc}, {Cho}, {Ekers}, {Day}, {Deller}, {Flynn}, {James}, {Macquart},
  {Mahony}, {Marnoch}, {Moss}, {Phillips}, {Prochaska}, {Qiu}, {Ryder},
  {Shannon}, {Tejos}, \& {Wong}}]{bhandari2020}
{Bhandari}, S., {Bannister}, K.~W., {Lenc}, E., {et~al.} 2020{\natexlab{a}},
  \apjl, 901, L20, \dodoi{10.3847/2041-8213/abb462}

\bibitem[{{Bhandari} {et~al.}(2020{\natexlab{b}}){Bhandari}, {Sadler},
  {Prochaska}, {Simha}, {Ryder}, {Marnoch}, {Bannister}, {Macquart}, {Flynn},
  {Shannon}, {Tejos}, {Corro-Guerra}, {Day}, {Deller}, {Ekers}, {Lopez},
  {Mahony}, {Nu{\~n}ez}, \& {Phillips}}]{bsp+20}
{Bhandari}, S., {Sadler}, E.~M., {Prochaska}, J.~X., {et~al.}
  2020{\natexlab{b}}, \apjl, 895, L37, \dodoi{10.3847/2041-8213/ab672e}

\bibitem[{{Bochenek} {et~al.}(2020{\natexlab{a}}){Bochenek}, {Ravi}, {Belov},
  {Hallinan}, {Kocz}, {Kulkarni}, \& {McKenna}}]{brb+20}
{Bochenek}, C.~D., {Ravi}, V., {Belov}, K.~V., {et~al.} 2020{\natexlab{a}},
  arXiv e-prints, arXiv:2005.10828.
\newblock \doarXiv{2005.10828}

\bibitem[{{Bochenek} {et~al.}(2020{\natexlab{b}}){Bochenek}, {Ravi}, \&
  {Dong}}]{bochenek2020}
{Bochenek}, C.~D., {Ravi}, V., \& {Dong}, D. 2020{\natexlab{b}}, arXiv
  e-prints, arXiv:2009.13030.
\newblock \doarXiv{2009.13030}

\bibitem[{{Bodaghee} {et~al.}(2012){Bodaghee}, {Tomsick}, {Rodriguez}, \&
  {James}}]{bodaghee2012}
{Bodaghee}, A., {Tomsick}, J.~A., {Rodriguez}, J., \& {James}, J.~B. 2012,
  \apj, 744, 108, \dodoi{10.1088/0004-637X/744/2/108}

\bibitem[{{Bozzo} {et~al.}(2008){Bozzo}, {Falanga}, \& {Stella}}]{bozzo2008}
{Bozzo}, E., {Falanga}, M., \& {Stella}, L. 2008, \apj, 683, 1031,
  \dodoi{10.1086/589990}

\bibitem[{{Carrasco} {et~al.}(2018){Carrasco}, {Gil de Paz}, {Gallego},
  {Iglesias-P{\'a}ramo}, {Cedazo}, {Garc{\'\i}a Vargas}, {Arrillaga},
  {Avil{\'e}s}, {Bouquin}, {Carbajo}, {Cardiel}, {Carrera}, {Castillo Morales},
  {Castillo-Dom{\'\i}nguez}, {Esteban San Rom{\'a}n}, {Ferrusca},
  {G{\'o}mez-{\'A}lvarez}, {Izazaga-P{\'e}rez}, {Lefort}, {L{\'o}pez Orozco},
  {Maldonado}, {Mart{\'\i}nez Delgado}, {Morales Dur{\'a}n}, {M{\'u}jica},
  {Ortiz}, {P{\'a}ez}, {Pascual}, {P{\'e}rez-Calpena}, {Picazo},
  {S{\'a}nchez-Penim}, {S{\'a}nchez-Blanco}, {Tulloch}, {Vel{\'a}zquez},
  {V{\'\i}lchez}, {Zamorano}, {Aguerri}, {Barrado}, {Bertone}, {Cava},
  {Catal{\'a}n-Torrecilla}, {Cenarro}, {Ch{\'a}vez}, {Dullo}, {Eliche},
  {Garc{\'\i}a}, {Garc{\'\i}a-Rojas}, {Guichard}, {Gonz{\'a}lez-Delgado},
  {Guzm{\'a}n}, {Herrero}, {Hu{\'e}lamo}, {Hughes}, {Jim{\'e}nez-Vicente},
  {Kehrig}, {Marino}, {M{\'a}rquez}, {Masegosa}, {Mayya}, {M{\'e}ndez-Abreu},
  {Moll{\'a}}, {Mu{\~n}oz-Tu{\~n}{\'o}n}, {Peimbert}, {P{\'e}rez-Gonz{\'a}lez},
  {P{\'e}rez-Montero}, {Roca-F{\`a}brega}, {Rodr{\'\i}guez},
  {Rodr{\'\i}guez-Espinosa}, {Rodr{\'\i}guez-Merino},
  {Rodr{\'\i}guez-Mu{\~n}oz}, {Rosa-Gonz{\'a}lez}, {S{\'a}nchez-Almeida},
  {S{\'a}nchez Contreras}, {S{\'a}nchez-Bl{\'a}zquez}, {S{\'a}nchez},
  {Sarajedini}, {Silich}, {Sim{\'o}n-D{\'\i}az}, {Tenorio-Tagle}, {Terlevich},
  {Terlevich}, {Torres-Peimbert}, {Trujillo}, {Tsamis}, \&
  {Vega}}]{carrasco2018}
{Carrasco}, E., {Gil de Paz}, A., {Gallego}, J., {et~al.} 2018, in Society of
  Photo-Optical Instrumentation Engineers (SPIE) Conference Series, Vol. 10702,
  Ground-based and Airborne Instrumentation for Astronomy VII, ed. C.~J.
  {Evans}, L.~{Simard}, \& H.~{Takami}, 1070216, \dodoi{10.1117/12.2313040}

\bibitem[{{Chatterjee} {et~al.}(2017){Chatterjee}, {Law}, {Wharton},
  {Burke-Spolaor}, {Hessels}, {Bower}, {Cordes}, {Tendulkar}, {Bassa},
  {Demorest}, {Butler}, {Seymour}, {Scholz}, {Abruzzo}, {Bogdanov}, {Kaspi},
  {Keimpema}, {Lazio}, {Marcote}, {McLaughlin}, {Paragi}, {Ransom}, {Rupen},
  {Spitler}, \& {van Langevelde}}]{clw+17}
{Chatterjee}, S., {Law}, C.~J., {Wharton}, R.~S., {et~al.} 2017, \nat, 541, 58.
\newblock \doarXiv{1701.01098}

\bibitem[{{CHIME/FRB Collaboration} {et~al.}(2019{\natexlab{a}}){CHIME/FRB
  Collaboration}, {Andersen}, {Band ura}, {Bhardwaj}, {Boubel}, {Boyce},
  {Boyle}, {Brar}, {Cassanelli}, {Chawla}, {Cubranic}, {Deng}, {Dobbs},
  {Fandino}, {Fonseca}, {Gaensler}, {Gilbert}, {Giri}, {Good}, {Halpern},
  {H{\"o}fer}, {Hill}, {Hinshaw}, {Josephy}, {Kaspi}, {Kothes}, {Landecker},
  {Lang}, {Li}, {Lin}, {Masui}, {Mena-Parra}, {Merryfield}, {Mckinven},
  {Michilli}, {Milutinovic}, {Naidu}, {Newburgh}, {Ng}, {Patel}, {Pen},
  {Pinsonneault-Marotte}, {Pleunis}, {Rafiei-Ravandi}, {Rahman}, {Ransom},
  {Renard}, {Scholz}, {Siegel}, {Singh}, {Smith}, {Stairs}, {Tendulkar},
  {Tretyakov}, {Vanderlinde}, {Yadav}, \& {Zwaniga}}]{abb+19c}
{CHIME/FRB Collaboration}, {Andersen}, B.~C., {Band ura}, K., {et~al.}
  2019{\natexlab{a}}, \apjl, 885, L24.
\newblock \doarXiv{1908.03507}

\bibitem[{{CHIME/FRB Collaboration} {et~al.}(2019{\natexlab{b}}){CHIME/FRB
  Collaboration}, {Amiri}, {Bandura}, {Bhardwaj}, {Boubel}, {Boyce}, {Boyle},
  {.~Brar}, {Burhanpurkar}, {Cassanelli}, {Chawla}, {Cliche}, {Cubranic},
  {Deng}, {Denman}, {Dobbs}, {Fandino}, {Fonseca}, {Gaensler}, {Gilbert},
  {Gill}, {Giri}, {Good}, {Halpern}, {Hanna}, {Hill}, {Hinshaw}, {H{\"o}fer},
  {Josephy}, {Kaspi}, {Landecker}, {Lang}, {Lin}, {Masui}, {Mckinven},
  {Mena-Parra}, {Merryfield}, {Michilli}, {Milutinovic}, {Moatti}, {Naidu},
  {Newburgh}, {Ng}, {Patel}, {Pen}, {Pinsonneault-Marotte}, {Pleunis},
  {Rafiei-Ravandi}, {Rahman}, {Ransom}, {Renard}, {Scholz}, {Shaw}, {Siegel},
  {Smith}, {Stairs}, {Tendulkar}, {Tretyakov}, {Vanderlinde}, \&
  {Yadav}}]{abb+19b}
{CHIME/FRB Collaboration}, {Amiri}, M., {Bandura}, K., {et~al.}
  2019{\natexlab{b}}, \nat, 566, 235, \dodoi{10.1038/s41586-018-0864-x}

\bibitem[{{CHIME/FRB Collaboration} {et~al.}(2020{\natexlab{a}}){CHIME/FRB
  Collaboration}, {Andersen}, {Bandura}, {Bhardwaj}, {Bij}, {Boyce}, {Boyle},
  {Brar}, {Cassanelli}, {Chawla}, {Chen}, {Cliche}, {Cook}, {Cubranic},
  {Curtin}, {Denman}, {Dobbs}, {Dong}, {Fandino}, {Fonseca}, {Gaensler},
  {Giri}, {Good}, {Halpern}, {Hill}, {Hinshaw}, {H{\"o}fer}, {Josephy},
  {Kania}, {Kaspi}, {Landecker}, {Leung}, {Li}, {Lin}, {Masui}, {Mckinven},
  {Mena-Parra}, {Merryfield}, {Meyers}, {Michilli}, {Milutinovic},
  {Mirhosseini}, {M{\"u}nchmeyer}, {Naidu}, {Newburgh}, {Ng}, {Patel}, {Pen},
  {Pinsonneault-Marotte}, {Pleunis}, {Quine}, {Rafiei-Ravandi}, {Rahman},
  {Ransom}, {Renard}, {Sanghavi}, {Scholz}, {Shaw}, {Shin}, {Siegel}, {Singh},
  {Smegal}, {Smith}, {Stairs}, {Tan}, {Tendulkar}, {Tretyakov}, {Vanderlinde},
  {Wang}, {Wulf}, \& {Zwaniga}}]{abb+20}
{CHIME/FRB Collaboration}, {Andersen}, B.~C., {Bandura}, K.~M., {et~al.}
  2020{\natexlab{a}}, arXiv e-prints, arXiv:2005.10324.
\newblock \doarXiv{2005.10324}

\bibitem[{{CHIME/FRB Collaboration} {et~al.}(2020{\natexlab{b}}){CHIME/FRB
  Collaboration}, {Amiri}, {Andersen}, {Band ura}, {Bhardwaj}, {Boyle}, {Brar},
  {Chawla}, {Chen}, {Cliche}, {Cubranic}, {Deng}, {Denman}, {Dobbs}, {Dong},
  {Fand ino}, {Fonseca}, {Gaensler}, {Giri}, {Good}, {Halpern}, {Hessels},
  {Hill}, {H{\"o}fer}, {Josephy}, {Kania}, {Karuppusamy}, {Kaspi}, {Keimpema},
  {Kirsten}, {Landecker}, {Lang}, {Leung}, {Li}, {Lin}, {Marcote}, {Masui},
  {McKinven}, {Mena-Parra}, {Merryfield}, {Michilli}, {Milutinovic},
  {Mirhosseini}, {Naidu}, {Newburgh}, {Ng}, {Nimmo}, {Paragi}, {Patel}, {Pen},
  {Pinsonneault-Marotte}, {Pleunis}, {Rafiei-Ravandi}, {Rahman}, {Ransom},
  {Renard}, {Sanghavi}, {Scholz}, {Shaw}, {Shin}, {Siegel}, {Singh}, {Smegal},
  {Smith}, {Stairs}, {Tendulkar}, {Tretyakov}, {Vanderlinde}, {Wang}, {Wang},
  {Wulf}, {Yadav}, \& {Zwaniga}}]{r3_period_2020}
{CHIME/FRB Collaboration}, {Amiri}, M., {Andersen}, B.~C., {et~al.}
  2020{\natexlab{b}}, \nat, 582, 351, \dodoi{10.1038/s41586-020-2398-2}

\bibitem[{{Chittidi} {et~al.}(2020){Chittidi}, {Simha}, {Mannings},
  {Prochaska}, {Rafelski}, {Neeleman}, {Macquart}, {Tejos}, {Jorgenson},
  {Ryder}, {Day}, {Marnoch}, {Bhandari}, {Deller}, {Qiu}, {Bannister},
  {Shannon}, \& {Heintz}}]{chittidi2020}
{Chittidi}, J.~S., {Simha}, S., {Mannings}, A., {et~al.} 2020, arXiv e-prints,
  arXiv:2005.13158.
\newblock \doarXiv{2005.13158}

\bibitem[{{Cordes} \& {Chatterjee}(2019)}]{cc19}
{Cordes}, J.~M., \& {Chatterjee}, S. 2019, \araa, 57, 417,
  \dodoi{10.1146/annurev-astro-091918-104501}

\bibitem[{{Cruces} {et~al.}(2020){Cruces}, {Spitler}, {Scholz}, {Lynch},
  {Seymour}, {Hessels}, {Gouiff{\`e}s}, {Hilmarsson}, {Kramer}, \&
  {Munjal}}]{cruces2020}
{Cruces}, M., {Spitler}, L.~G., {Scholz}, P., {et~al.} 2020, arXiv e-prints,
  arXiv:2008.03461.
\newblock \doarXiv{2008.03461}

\bibitem[{{Cui} {et~al.}(2021){Cui}, {Zhang}, {Wang}, {Zhang}, {Li}, {Peng},
  {Zhu}, {Wang}, {Strom}, {Ye}, {Wang}, \& {Yang}}]{cui2021}
{Cui}, X.-H., {Zhang}, C.-M., {Wang}, S.-Q., {et~al.} 2021, \mnras, 500, 3275,
  \dodoi{10.1093/mnras/staa3351}

\bibitem[{{Day} {et~al.}(2020){Day}, {Deller}, {Shannon}, {Qiu}, {Bannister},
  {Bhandari}, {Ekers}, {Flynn}, {James}, {Macquart}, {Mahony}, {Phillips}, \&
  {Xavier Prochaska}}]{day2020}
{Day}, C.~K., {Deller}, A.~T., {Shannon}, R.~M., {et~al.} 2020, \mnras, 497,
  3335, \dodoi{10.1093/mnras/staa2138}

\bibitem[{{Ding} {et~al.}(2020){Ding}, {Deller}, {Lower}, {Flynn},
  {Chatterjee}, {Brisken}, {Hurley-Walker}, {Camilo}, {Sarkissian}, \&
  {Gupta}}]{ding2020}
{Ding}, H., {Deller}, A.~T., {Lower}, M.~E., {et~al.} 2020, \mnras, 498, 3736,
  \dodoi{10.1093/mnras/staa2531}

\bibitem[{{Dolphin}(2016)}]{dolphot2016}
{Dolphin}, A. 2016, {DOLPHOT: Stellar photometry}.
\newblock \doeprint{1608.013}

\bibitem[{{Dubus}(2013)}]{dubus2013}
{Dubus}, G. 2013, \aapr, 21, 64, \dodoi{10.1007/s00159-013-0064-5}

\bibitem[{{Farah} {et~al.}(2018){Farah}, {Flynn}, {Bailes}, {Jameson},
  {Bannister}, {Barr}, {Bateman}, {Bhandari}, {Caleb}, {Campbell-Wilson},
  {Chang}, {Deller}, {Green}, {Hunstead}, {Jankowski}, {Keane}, {Macquart},
  {M{\"o}ller}, {Onken}, {Os{\l}owski}, {Parthasarathy}, {Plant}, {Ravi},
  {Shannon}, {Tucker}, {Venkatraman Krishnan}, \& {Wolf}}]{ffb18}
{Farah}, W., {Flynn}, C., {Bailes}, M., {et~al.} 2018, \mnras, 478, 1209,
  \dodoi{10.1093/mnras/sty1122}

\bibitem[{{Fich} {et~al.}(1990){Fich}, {Treffers}, \& {Dahl}}]{fich1990}
{Fich}, M., {Treffers}, R.~R., \& {Dahl}, G.~P. 1990, \aj, 99, 622,
  \dodoi{10.1086/115356}

\bibitem[{{Fonseca} {et~al.}(2020){Fonseca}, {Andersen}, {Bhardwaj}, {Chawla},
  {Good}, {Josephy}, {Kaspi}, {Masui}, {Mckinven}, {Michilli}, {Pleunis},
  {Shin}, {Tendulkar}, {Bandura}, {Boyle}, {Brar}, {Cassanelli}, {Cubranic},
  {Dobbs}, {Dong}, {Gaensler}, {Hinshaw}, {Land ecker}, {Leung}, {Li}, {Lin},
  {Mena-Parra}, {Merryfield}, {Naidu}, {Ng}, {Patel}, {Pen}, {Rafiei-Ravandi},
  {Rahman}, {Ransom}, {Scholz}, {Smith}, {Stairs}, {Vanderlinde}, {Yadav}, \&
  {Zwaniga}}]{fab+20}
{Fonseca}, E., {Andersen}, B.~C., {Bhardwaj}, M., {et~al.} 2020, \apjl, 891,
  L6, \dodoi{10.3847/2041-8213/ab7208}

\bibitem[{{Fujii} \& {Portegies Zwart}(2011)}]{fujii2011}
{Fujii}, M.~S., \& {Portegies Zwart}, S. 2011, Science, 334, 1380,
  \dodoi{10.1126/science.1211927}

\bibitem[{{Gaia Collaboration} {et~al.}(2016){Gaia Collaboration}, {Prusti},
  {de Bruijne}, {Brown}, {Vallenari}, {Babusiaux}, {Bailer-Jones}, {Bastian},
  {Biermann}, {Evans}, {Eyer}, {Jansen}, {Jordi}, {Klioner}, {Lammers},
  {Lindegren}, {Luri}, {Mignard}, {Milligan}, {Panem}, {Poinsignon},
  {Pourbaix}, {Randich}, {Sarri}, {Sartoretti}, {Siddiqui}, {Soubiran},
  {Valette}, {van Leeuwen}, {Walton}, {Aerts}, {Arenou}, {Cropper}, {Drimmel},
  {H{\o}g}, {Katz}, {Lattanzi}, {O'Mullane}, {Grebel}, {Holland}, {Huc},
  {Passot}, {Bramante}, {Cacciari}, {Casta{\~n}eda}, {Chaoul}, {Cheek}, {De
  Angeli}, {Fabricius}, {Guerra}, {Hern{\'a}ndez}, {Jean-Antoine-Piccolo},
  {Masana}, {Messineo}, {Mowlavi}, {Nienartowicz}, {Ord{\'o}{\~n}ez-Blanco},
  {Panuzzo}, {Portell}, {Richards}, {Riello}, {Seabroke}, {Tanga},
  {Th{\'e}venin}, {Torra}, {Els}, {Gracia-Abril}, {Comoretto},
  {Garcia-Reinaldos}, {Lock}, {Mercier}, {Altmann}, {Andrae}, {Astraatmadja},
  {Bellas-Velidis}, {Benson}, {Berthier}, {Blomme}, {Busso}, {Carry},
  {Cellino}, {Clementini}, {Cowell}, {Creevey}, {Cuypers}, {Davidson}, {De
  Ridder}, {de Torres}, {Delchambre}, {Dell'Oro}, {Ducourant}, {Fr{\'e}mat},
  {Garc{\'\i}a-Torres}, {Gosset}, {Halbwachs}, {Hambly}, {Harrison}, {Hauser},
  {Hestroffer}, {Hodgkin}, {Huckle}, {Hutton}, {Jasniewicz}, {Jordan},
  {Kontizas}, {Korn}, {Lanzafame}, {Manteiga}, {Moitinho}, {Muinonen},
  {Osinde}, {Pancino}, {Pauwels}, {Petit}, {Recio-Blanco}, {Robin}, {Sarro},
  {Siopis}, {Smith}, {Smith}, {Sozzetti}, {Thuillot}, {van Reeven}, {Viala},
  {Abbas}, {Abreu Aramburu}, {Accart}, {Aguado}, {Allan}, {Allasia},
  {Altavilla}, {{\'A}lvarez}, {Alves}, {Anderson}, {Andrei}, {Anglada Varela},
  {Antiche}, {Antoja}, {Ant{\'o}n}, {Arcay}, {Atzei}, {Ayache}, {Bach},
  {Baker}, {Balaguer-N{\'u}{\~n}ez}, {Barache}, {Barata}, {Barbier}, {Barblan},
  {Baroni}, {Barrado y Navascu{\'e}s}, {Barros}, {Barstow}, {Becciani},
  {Bellazzini}, {Bellei}, {Bello Garc{\'\i}a}, {Belokurov}, {Bendjoya},
  {Berihuete}, {Bianchi}, {Bienaym{\'e}}, {Billebaud}, {Blagorodnova},
  {Blanco-Cuaresma}, {Boch}, {Bombrun}, {Borrachero}, {Bouquillon}, {Bourda},
  {Bouy}, {Bragaglia}, {Breddels}, {Brouillet}, {Br{\"u}semeister},
  {Bucciarelli}, {Budnik}, {Burgess}, {Burgon}, {Burlacu}, {Busonero}, {Buzzi},
  {Caffau}, {Cambras}, {Campbell}, {Cancelliere}, {Cantat-Gaudin}, {Carlucci},
  {Carrasco}, {Castellani}, {Charlot}, {Charnas}, {Charvet}, {Chassat},
  {Chiavassa}, {Clotet}, {Cocozza}, {Collins}, {Collins}, {Costigan}, {Crifo},
  {Cross}, {Crosta}, {Crowley}, {Dafonte}, {Damerdji}, {Dapergolas}, {David},
  {David}, {De Cat}, {de Felice}, {de Laverny}, {De Luise}, {De March}, {de
  Martino}, {de Souza}, {Debosscher}, {del Pozo}, {Delbo}, {Delgado},
  {Delgado}, {di Marco}, {Di Matteo}, {Diakite}, {Distefano}, {Dolding}, {Dos
  Anjos}, {Drazinos}, {Dur{\'a}n}, {Dzigan}, {Ecale}, {Edvardsson}, {Enke},
  {Erdmann}, {Escolar}, {Espina}, {Evans}, {Eynard Bontemps}, {Fabre},
  {Fabrizio}, {Faigler}, {Falc{\~a}o}, {Farr{\`a}s Casas}, {Faye}, {Federici},
  {Fedorets}, {Fern{\'a}ndez-Hern{\'a}ndez}, {Fernique}, {Fienga}, {Figueras},
  {Filippi}, {Findeisen}, {Fonti}, {Fouesneau}, {Fraile}, {Fraser}, {Fuchs},
  {Furnell}, {Gai}, {Galleti}, {Galluccio}, {Garabato}, {Garc{\'\i}a-Sedano},
  {Gar{\'e}}, {Garofalo}, {Garralda}, {Gavras}, {Gerssen}, {Geyer}, {Gilmore},
  {Girona}, {Giuffrida}, {Gomes}, {Gonz{\'a}lez-Marcos},
  {Gonz{\'a}lez-N{\'u}{\~n}ez}, {Gonz{\'a}lez-Vidal}, {Granvik}, {Guerrier},
  {Guillout}, {Guiraud}, {G{\'u}rpide}, {Guti{\'e}rrez-S{\'a}nchez}, {Guy},
  {Haigron}, {Hatzidimitriou}, {Haywood}, {Heiter}, {Helmi}, {Hobbs},
  {Hofmann}, {Holl}, {Holland }, {Hunt}, {Hypki}, {Icardi}, {Irwin}, {Jevardat
  de Fombelle}, {Jofr{\'e}}, {Jonker}, {Jorissen}, {Julbe}, {Karampelas},
  {Kochoska}, {Kohley}, {Kolenberg}, {Kontizas}, {Koposov}, {Kordopatis},
  {Koubsky}, {Kowalczyk}, {Krone-Martins}, {Kudryashova}, {Kull}, {Bachchan},
  {Lacoste-Seris}, {Lanza}, {Lavigne}, {Le Poncin-Lafitte}, {Lebreton},
  {Lebzelter}, {Leccia}, {Leclerc}, {Lecoeur-Taibi}, {Lemaitre}, {Lenhardt},
  {Leroux}, {Liao}, {Licata}, {Lindstr{\o}m}, {Lister}, {Livanou}, {Lobel},
  {L{\"o}ffler}, {L{\'o}pez}, {Lopez-Lozano}, {Lorenz}, {Loureiro},
  {MacDonald}, {Magalh{\~a}es Fernandes}, {Managau}, {Mann}, {Mantelet},
  {Marchal}, {Marchant}, {Marconi}, {Marie}, {Marinoni}, {Marrese},
  {Marschalk{\'o}}, {Marshall}, {Mart{\'\i}n-Fleitas}, {Martino}, {Mary},
  {Matijevi{\v{c}}}, {Mazeh}, {McMillan}, {Messina}, {Mestre}, {Michalik},
  {Millar}, {Miranda}, {Molina}, {Molinaro}, {Molinaro}, {Moln{\'a}r},
  {Moniez}, {Montegriffo}, {Monteiro}, {Mor}, {Mora}, {Morbidelli}, {Morel},
  {Morgenthaler}, {Morley}, {Morris}, {Mulone}, {Muraveva}, {Musella},
  {Narbonne}, {Nelemans}, {Nicastro}, {Noval}, {Ord{\'e}novic},
  {Ordieres-Mer{\'e}}, {Osborne}, {Pagani}, {Pagano}, {Pailler}, {Palacin},
  {Palaversa}, {Parsons}, {Paulsen}, {Pecoraro}, {Pedrosa}, {Pentik{\"a}inen},
  {Pereira}, {Pichon}, {Piersimoni}, {Pineau}, {Plachy}, {Plum}, {Poujoulet},
  {Pr{\v{s}}a}, {Pulone}, {Ragaini}, {Rago}, {Rambaux}, {Ramos-Lerate},
  {Ranalli}, {Rauw}, {Read}, {Regibo}, {Renk}, {Reyl{\'e}}, {Ribeiro},
  {Rimoldini}, {Ripepi}, {Riva}, {Rixon}, {Roelens}, {Romero-G{\'o}mez},
  {Rowell}, {Royer}, {Rudolph}, {Ruiz-Dern}, {Sadowski}, {Sagrist{\`a}
  Sell{\'e}s}, {Sahlmann}, {Salgado}, {Salguero}, {Sarasso}, {Savietto},
  {Schnorhk}, {Schultheis}, {Sciacca}, {Segol}, {Segovia}, {Segransan},
  {Serpell}, {Shih}, {Smareglia}, {Smart}, {Smith}, {Solano}, {Solitro},
  {Sordo}, {Soria Nieto}, {Souchay}, {Spagna}, {Spoto}, {Stampa}, {Steele},
  {Steidelm{\"u}ller}, {Stephenson}, {Stoev}, {Suess}, {S{\"u}veges}, {Surdej},
  {Szabados}, {Szegedi-Elek}, {Tapiador}, {Taris}, {Tauran}, {Taylor},
  {Teixeira}, {Terrett}, {Tingley}, {Trager}, {Turon}, {Ulla}, {Utrilla},
  {Valentini}, {van Elteren}, {Van Hemelryck}, {van Leeuwen}, {Varadi},
  {Vecchiato}, {Veljanoski}, {Via}, {Vicente}, {Vogt}, {Voss}, {Votruba},
  {Voutsinas}, {Walmsley}, {Weiler}, {Weingrill}, {Werner}, {Wevers},
  {Whitehead}, {Wyrzykowski}, {Yoldas}, {{\v{Z}}erjal}, {Zucker}, {Zurbach},
  {Zwitter}, {Alecu}, {Allen}, {Allende Prieto}, {Amorim},
  {Anglada-Escud{\'e}}, {Arsenijevic}, {Azaz}, {Balm}, {Beck}, {Bernstein},
  {Bigot}, {Bijaoui}, {Blasco}, {Bonfigli}, {Bono}, {Boudreault}, {Bressan},
  {Brown}, {Brunet}, {Bunclark}, {Buonanno}, {Butkevich}, {Carret}, {Carrion},
  {Chemin}, {Ch{\'e}reau}, {Corcione}, {Darmigny}, {de Boer}, {de Teodoro}, {de
  Zeeuw}, {Delle Luche}, {Domingues}, {Dubath}, {Fodor}, {Fr{\'e}zouls},
  {Fries}, {Fustes}, {Fyfe}, {Gallardo}, {Gallegos}, {Gardiol}, {Gebran},
  {Gomboc}, {G{\'o}mez}, {Grux}, {Gueguen}, {Heyrovsky}, {Hoar}, {Iannicola},
  {Isasi Parache}, {Janotto}, {Joliet}, {Jonckheere}, {Keil}, {Kim},
  {Klagyivik}, {Klar}, {Knude}, {Kochukhov}, {Kolka}, {Kos}, {Kutka}, {Lainey},
  {LeBouquin}, {Liu}, {Loreggia}, {Makarov}, {Marseille}, {Martayan},
  {Martinez-Rubi}, {Massart}, {Meynadier}, {Mignot}, {Munari}, {Nguyen},
  {Nordlander}, {Ocvirk}, {O'Flaherty}, {Olias Sanz}, {Ortiz}, {Osorio},
  {Oszkiewicz}, {Ouzounis}, {Palmer}, {Park}, {Pasquato}, {Peltzer}, {Peralta},
  {P{\'e}turaud}, {Pieniluoma}, {Pigozzi}, {Poels}, {Prat}, {Prod'homme},
  {Raison}, {Rebordao}, {Risquez}, {Rocca-Volmerange}, {Rosen}, {Ruiz-Fuertes},
  {Russo}, {Sembay}, {Serraller Vizcaino}, {Short}, {Siebert}, {Silva},
  {Sinachopoulos}, {Slezak}, {Soffel}, {Sosnowska}, {Strai{\v{z}}ys}, {ter
  Linden}, {Terrell}, {Theil}, {Tiede}, {Troisi}, {Tsalmantza}, {Tur},
  {Vaccari}, {Vachier}, {Valles}, {Van Hamme}, {Veltz}, {Virtanen}, {Wallut},
  {Wichmann}, {Wilkinson}, {Ziaeepour}, \& {Zschocke}}]{gaiamission}
{Gaia Collaboration}, {Prusti}, T., {de Bruijne}, J.~H.~J., {et~al.} 2016,
  \aap, 595, A1, \dodoi{10.1051/0004-6361/201629272}

\bibitem[{{Gaia Collaboration} {et~al.}(2018){Gaia Collaboration}, {Brown},
  {Vallenari}, {Prusti}, {de Bruijne}, {Babusiaux}, {Bailer-Jones}, {Biermann},
  {Evans}, {Eyer}, {Jansen}, {Jordi}, {Klioner}, {Lammers}, {Lindegren},
  {Luri}, {Mignard}, {Panem}, {Pourbaix}, {Randich}, {Sartoretti}, {Siddiqui},
  {Soubiran}, {van Leeuwen}, {Walton}, {Arenou}, {Bastian}, {Cropper},
  {Drimmel}, {Katz}, {Lattanzi}, {Bakker}, {Cacciari}, {Casta{\~n}eda},
  {Chaoul}, {Cheek}, {De Angeli}, {Fabricius}, {Guerra}, {Holl}, {Masana},
  {Messineo}, {Mowlavi}, {Nienartowicz}, {Panuzzo}, {Portell}, {Riello},
  {Seabroke}, {Tanga}, {Th{\'e}venin}, {Gracia-Abril}, {Comoretto},
  {Garcia-Reinaldos}, {Teyssier}, {Altmann}, {Andrae}, {Audard},
  {Bellas-Velidis}, {Benson}, {Berthier}, {Blomme}, {Burgess}, {Busso},
  {Carry}, {Cellino}, {Clementini}, {Clotet}, {Creevey}, {Davidson}, {De
  Ridder}, {Delchambre}, {Dell'Oro}, {Ducourant},
  {Fern{\'a}ndez-Hern{\'a}ndez}, {Fouesneau}, {Fr{\'e}mat}, {Galluccio},
  {Garc{\'\i}a-Torres}, {Gonz{\'a}lez-N{\'u}{\~n}ez}, {Gonz{\'a}lez-Vidal},
  {Gosset}, {Guy}, {Halbwachs}, {Hambly}, {Harrison}, {Hern{\'a}ndez},
  {Hestroffer}, {Hodgkin}, {Hutton}, {Jasniewicz}, {Jean-Antoine-Piccolo},
  {Jordan}, {Korn}, {Krone-Martins}, {Lanzafame}, {Lebzelter}, {L{\"o}ffler},
  {Manteiga}, {Marrese}, {Mart{\'\i}n-Fleitas}, {Moitinho}, {Mora}, {Muinonen},
  {Osinde}, {Pancino}, {Pauwels}, {Petit}, {Recio-Blanco}, {Richards},
  {Rimoldini}, {Robin}, {Sarro}, {Siopis}, {Smith}, {Sozzetti}, {S{\"u}veges},
  {Torra}, {van Reeven}, {Abbas}, {Abreu Aramburu}, {Accart}, {Aerts},
  {Altavilla}, {{\'A}lvarez}, {Alvarez}, {Alves}, {Anderson}, {Andrei},
  {Anglada Varela}, {Antiche}, {Antoja}, {Arcay}, {Astraatmadja}, {Bach},
  {Baker}, {Balaguer-N{\'u}{\~n}ez}, {Balm}, {Barache}, {Barata}, {Barbato},
  {Barblan}, {Barklem}, {Barrado}, {Barros}, {Barstow}, {Bartholom{\'e}
  Mu{\~n}oz}, {Bassilana}, {Becciani}, {Bellazzini}, {Berihuete}, {Bertone},
  {Bianchi}, {Bienaym{\'e}}, {Blanco-Cuaresma}, {Boch}, {Boeche}, {Bombrun},
  {Borrachero}, {Bossini}, {Bouquillon}, {Bourda}, {Bragaglia}, {Bramante},
  {Breddels}, {Bressan}, {Brouillet}, {Br{\"u}semeister}, {Brugaletta},
  {Bucciarelli}, {Burlacu}, {Busonero}, {Butkevich}, {Buzzi}, {Caffau},
  {Cancelliere}, {Cannizzaro}, {Cantat-Gaudin}, {Carballo}, {Carlucci},
  {Carrasco}, {Casamiquela}, {Castellani}, {Castro-Ginard}, {Charlot},
  {Chemin}, {Chiavassa}, {Cocozza}, {Costigan}, {Cowell}, {Crifo}, {Crosta},
  {Crowley}, {Cuypers}, {Dafonte}, {Damerdji}, {Dapergolas}, {David}, {David},
  {de Laverny}, {De Luise}, {De March}, {de Martino}, {de Souza}, {de Torres},
  {Debosscher}, {del Pozo}, {Delbo}, {Delgado}, {Delgado}, {Di Matteo},
  {Diakite}, {Diener}, {Distefano}, {Dolding}, {Drazinos}, {Dur{\'a}n},
  {Edvardsson}, {Enke}, {Eriksson}, {Esquej}, {Eynard Bontemps}, {Fabre},
  {Fabrizio}, {Faigler}, {Falc{\~a}o}, {Farr{\`a}s Casas}, {Federici},
  {Fedorets}, {Fernique}, {Figueras}, {Filippi}, {Findeisen}, {Fonti},
  {Fraile}, {Fraser}, {Fr{\'e}zouls}, {Gai}, {Galleti}, {Garabato},
  {Garc{\'\i}a-Sedano}, {Garofalo}, {Garralda}, {Gavel}, {Gavras}, {Gerssen},
  {Geyer}, {Giacobbe}, {Gilmore}, {Girona}, {Giuffrida}, {Glass}, {Gomes},
  {Granvik}, {Gueguen}, {Guerrier}, {Guiraud}, {Guti{\'e}rrez-S{\'a}nchez},
  {Haigron}, {Hatzidimitriou}, {Hauser}, {Haywood}, {Heiter}, {Helmi}, {Heu},
  {Hilger}, {Hobbs}, {Hofmann}, {Holland}, {Huckle}, {Hypki}, {Icardi},
  {Jan{\ss}en}, {Jevardat de Fombelle}, {Jonker}, {Juh{\'a}sz}, {Julbe},
  {Karampelas}, {Kewley}, {Klar}, {Kochoska}, {Kohley}, {Kolenberg},
  {Kontizas}, {Kontizas}, {Koposov}, {Kordopatis}, {Kostrzewa-Rutkowska},
  {Koubsky}, {Lambert}, {Lanza}, {Lasne}, {Lavigne}, {Le Fustec}, {Le
  Poncin-Lafitte}, {Lebreton}, {Leccia}, {Leclerc}, {Lecoeur-Taibi},
  {Lenhardt}, {Leroux}, {Liao}, {Licata}, {Lindstr{\o}m}, {Lister}, {Livanou},
  {Lobel}, {L{\'o}pez}, {Managau}, {Mann}, {Mantelet}, {Marchal}, {Marchant},
  {Marconi}, {Marinoni}, {Marschalk{\'o}}, {Marshall}, {Martino}, {Marton},
  {Mary}, {Massari}, {Matijevi{\v{c}}}, {Mazeh}, {McMillan}, {Messina},
  {Michalik}, {Millar}, {Molina}, {Molinaro}, {Moln{\'a}r}, {Montegriffo},
  {Mor}, {Morbidelli}, {Morel}, {Morris}, {Mulone}, {Muraveva}, {Musella},
  {Nelemans}, {Nicastro}, {Noval}, {O'Mullane}, {Ord{\'e}novic},
  {Ord{\'o}{\~n}ez-Blanco}, {Osborne}, {Pagani}, {Pagano}, {Pailler},
  {Palacin}, {Palaversa}, {Panahi}, {Pawlak}, {Piersimoni}, {Pineau}, {Plachy},
  {Plum}, {Poggio}, {Poujoulet}, {Pr{\v{s}}a}, {Pulone}, {Racero}, {Ragaini},
  {Rambaux}, {Ramos-Lerate}, {Regibo}, {Reyl{\'e}}, {Riclet}, {Ripepi}, {Riva},
  {Rivard}, {Rixon}, {Roegiers}, {Roelens}, {Romero-G{\'o}mez}, {Rowell},
  {Royer}, {Ruiz-Dern}, {Sadowski}, {Sagrist{\`a} Sell{\'e}s}, {Sahlmann},
  {Salgado}, {Salguero}, {Sanna}, {Santana-Ros}, {Sarasso}, {Savietto},
  {Schultheis}, {Sciacca}, {Segol}, {Segovia}, {S{\'e}gransan}, {Shih},
  {Siltala}, {Silva}, {Smart}, {Smith}, {Solano}, {Solitro}, {Sordo}, {Soria
  Nieto}, {Souchay}, {Spagna}, {Spoto}, {Stampa}, {Steele},
  {Steidelm{\"u}ller}, {Stephenson}, {Stoev}, {Suess}, {Surdej}, {Szabados},
  {Szegedi-Elek}, {Tapiador}, {Taris}, {Tauran}, {Taylor}, {Teixeira},
  {Terrett}, {Teyssand ier}, {Thuillot}, {Titarenko}, {Torra Clotet}, {Turon},
  {Ulla}, {Utrilla}, {Uzzi}, {Vaillant}, {Valentini}, {Valette}, {van Elteren},
  {Van Hemelryck}, {van Leeuwen}, {Vaschetto}, {Vecchiato}, {Veljanoski},
  {Viala}, {Vicente}, {Vogt}, {von Essen}, {Voss}, {Votruba}, {Voutsinas},
  {Walmsley}, {Weiler}, {Wertz}, {Wevers}, {Wyrzykowski}, {Yoldas},
  {{\v{Z}}erjal}, {Ziaeepour}, {Zorec}, {Zschocke}, {Zucker}, {Zurbach}, \&
  {Zwitter}}]{gaiadr2}
{Gaia Collaboration}, {Brown}, A.~G.~A., {Vallenari}, A., {et~al.} 2018, \aap,
  616, A1, \dodoi{10.1051/0004-6361/201833051}

\bibitem[{{Gajjar} {et~al.}(2018){Gajjar}, {Siemion}, {Price}, {Law},
  {Michilli}, {Hessels}, {Chatterjee}, {Archibald}, {Bower}, {Brinkman},
  {Burke-Spolaor}, {Cordes}, {Croft}, {Enriquez}, {Foster}, {Gizani},
  {Hellbourg}, {Isaacson}, {Kaspi}, {Lazio}, {Lebofsky}, {Lynch}, {MacMahon},
  {McLaughlin}, {Ransom}, {Scholz}, {Seymour}, {Spitler}, {Tendulkar},
  {Werthimer}, \& {Zhang}}]{gsp+18}
{Gajjar}, V., {Siemion}, A.~P.~V., {Price}, D.~C., {et~al.} 2018, \apj, 863, 2,
  \dodoi{10.3847/1538-4357/aad005}

\bibitem[{{Gil de Paz} {et~al.}(2018){Gil de Paz}, {Carrasco}, {Gallego},
  {Iglesias-P{\'a}ramo}, {Cedazo}, {Garc{\'\i}a-Vargas}, {Arrillaga},
  {Avil{\'e}s}, {Bouquin}, {Carbajo}, {Cardiel}, {Carrera}, {Castillo-Morales},
  {Castillo-Dom{\'\i}nguez}, {Esteban San Rom{\'a}n}, {Ferrusca},
  {G{\'o}mez-{\'A}lvarez}, {Izazaga-P{\'e}rez}, {Lefort}, {L{\'o}pez-Orozco},
  {Maldonado}, {Mart{\'\i}nez-Delgado}, {Morales-Dur{\'a}n}, {Mujica},
  {P{\'a}ez}, {Pascual}, {P{\'e}rez-Calpena}, {Picazo}, {S{\'a}nchez-Penim},
  {S{\'a}nchez-Blanco}, {Tulloch}, {Vel{\'a}zquez}, {V{\'\i}lchez}, {Zamorano},
  {Aguerri}, {Barrado y Navascues}, {Berlanas}, {Bertone}, {Cava},
  {Catal{\'a}n-Torrecilla}, {Cenarro}, {Ch{\'a}vez}, {Dullo}, {Garc{\'\i}a},
  {Garc{\'\i}a-Rojas}, {Guichard}, {Gonz{\'a}lez-Delgado}, {Guzm{\'a}n},
  {Herrero}, {Hu{\'e}lamo}, {Hughes}, {Jim{\'e}nez-Vicente}, {Kehrig},
  {Marino}, {M{\'a}rquez}, {Masegosa}, {Mayya}, {M{\'e}ndez-Abreu},
  {Moll{\'a}}, {Mu{\~n}oz-Tu{\~n}{\'o}n}, {Peimbert}, {P{\'e}rez-Gonz{\'a}lez},
  {P{\'e}rez-Montero}, {Rodr{\'\i}guez}, {Rodr{\'\i}guez-Espinosa},
  {Rodr{\'\i}guez Merino}, {Rodr{\'\i}guez-Mu{\~n}oz}, {Rosa-Gonz{\'a}lez},
  {S{\'a}nchez-Almeida}, {S{\'a}nchez-Contreras}, {S{\'a}nchez-Bl{\'a}zquez},
  {S{\'a}nchez}, {Sarajedini}, {Silich}, {Sim{\'o}n-D{\'\i}az},
  {Tenorio-Tagle}, {Terlevich}, {Terlevich}, {Torres-Peimbert}, {Trujillo},
  {Tsamis}, \& {Vega}}]{gildepaz2018}
{Gil de Paz}, A., {Carrasco}, E., {Gallego}, J., {et~al.} 2018, in Society of
  Photo-Optical Instrumentation Engineers (SPIE) Conference Series, Vol. 10702,
  Ground-based and Airborne Instrumentation for Astronomy VII, ed. C.~J.
  {Evans}, L.~{Simard}, \& H.~{Takami}, 1070217, \dodoi{10.1117/12.2313299}

\bibitem[{{Green} {et~al.}(2019){Green}, {Schlafly}, {Zucker}, {Speagle}, \&
  {Finkbeiner}}]{green2019}
{Green}, G.~M., {Schlafly}, E., {Zucker}, C., {Speagle}, J.~S., \&
  {Finkbeiner}, D. 2019, \apj, 887, 93, \dodoi{10.3847/1538-4357/ab5362}

\bibitem[{{Guti{\'e}rrez} {et~al.}(2011){Guti{\'e}rrez}, {Beckman}, \&
  {Buenrostro}}]{gutierrez2011properties}
{Guti{\'e}rrez}, L., {Beckman}, J.~E., \& {Buenrostro}, V. 2011, \aj, 141, 113,
  \dodoi{10.1088/0004-6256/141/4/113}

\bibitem[{Harris {et~al.}(2020)Harris, Millman, van~der Walt, Gommers,
  Virtanen, Cournapeau, Wieser, Taylor, Berg, Smith, Kern, Picus, Hoyer, van
  Kerkwijk, Brett, Haldane, Fernández~del Río, Wiebe, Peterson,
  Gérard-Marchant, Sheppard, Reddy, Weckesser, Abbasi, Gohlke, \&
  Oliphant}]{2020NumPy-Array}
Harris, C.~R., Millman, K.~J., van~der Walt, S.~J., {et~al.} 2020, Nature, 585,
  357–362, \dodoi{10.1038/s41586-020-2649-2}

\bibitem[{{Heintz} {et~al.}(2020){Heintz}, {Prochaska}, {Simha}, {Platts},
  {Fong}, {Tejos}, {Ryder}, {Aggarwal}, {Bhandari}, {Day}, {Deller},
  {Kilpatrick}, {Law}, {Macquart}, {Mannings}, {Marnoch}, {Sadler}, \&
  {Shannon}}]{heintz2020}
{Heintz}, K.~E., {Prochaska}, J.~X., {Simha}, S., {et~al.} 2020, arXiv
  e-prints, arXiv:2009.10747.
\newblock \doarXiv{2009.10747}

\bibitem[{{Hessels} {et~al.}(2019){Hessels}, {Spitler}, {Seymour}, {Cordes},
  {Michilli}, {Lynch}, {Gourdji}, {Archibald}, {Bassa}, {Bower}, {Chatterjee},
  {Connor}, {Crawford}, {Deneva}, {Gajjar}, {Kaspi}, {Keimpema}, {Law},
  {Marcote}, {McLaughlin}, {Paragi}, {Petroff}, {Ransom}, {Scholz}, {Stappers},
  \& {Tendulkar}}]{hss+19}
{Hessels}, J.~W.~T., {Spitler}, L.~G., {Seymour}, A.~D., {et~al.} 2019, \apjl,
  876, L23, \dodoi{10.3847/2041-8213/ab13ae}

\bibitem[{{Hilmarsson} {et~al.}(2020){Hilmarsson}, {Michilli}, {Spitler},
  {Wharton}, {Demorest}, {Desvignes}, {Gourdji}, {Hackstein}, {Hessels},
  {Nimmo}, {Seymour}, {Kramer}, \& {McKinven}}]{hilmarsson2020b}
{Hilmarsson}, G.~H., {Michilli}, D., {Spitler}, L.~G., {et~al.} 2020, arXiv
  e-prints, arXiv:2009.12135.
\newblock \doarXiv{2009.12135}

\bibitem[{{Hobbs} {et~al.}(2005){Hobbs}, {Lorimer}, {Lyne}, \&
  {Kramer}}]{hobbs2005}
{Hobbs}, G., {Lorimer}, D.~R., {Lyne}, A.~G., \& {Kramer}, M. 2005, \mnras,
  360, 974, \dodoi{10.1111/j.1365-2966.2005.09087.x}

\bibitem[{{Ioka} \& {Zhang}(2020{\natexlab{a}})}]{iokazhang2020}
{Ioka}, K., \& {Zhang}, B. 2020{\natexlab{a}}, \apjl, 893, L26,
  \dodoi{10.3847/2041-8213/ab83fb}

\bibitem[{{Ioka} \& {Zhang}(2020{\natexlab{b}})}]{iz20}
---. 2020{\natexlab{b}}, arXiv e-prints, arXiv:2002.08297.
\newblock \doarXiv{2002.08297}

\bibitem[{{Jedrzejewski}(1987)}]{1987MNRAS.226..747J}
{Jedrzejewski}, R.~I. 1987, \mnras, 226, 747, \dodoi{10.1093/mnras/226.4.747}

\bibitem[{{Josephy} {et~al.}(2019){Josephy}, {Chawla}, {Fonseca}, {Ng},
  {Patel}, {Pleunis}, {Scholz}, {Andersen}, {Bandura}, \& {Bhardwaj}}]{jcf+19}
{Josephy}, A., {Chawla}, P., {Fonseca}, E., {et~al.} 2019, arXiv e-prints,
  arXiv:1906.11305.
\newblock \doarXiv{1906.11305}

\bibitem[{Kaspi \& Beloborodov(2017)}]{kaspi2017magnetars}
Kaspi, V.~M., \& Beloborodov, A.~M. 2017, Annual Review of Astronomy and
  Astrophysics, 55, 261

\bibitem[{{Kewley} {et~al.}(2002){Kewley}, {Geller}, {Jansen}, \&
  {Dopita}}]{kewley2002}
{Kewley}, L.~J., {Geller}, M.~J., {Jansen}, R.~A., \& {Dopita}, M.~A. 2002,
  \aj, 124, 3135, \dodoi{10.1086/344487}

\bibitem[{{Kirsten} {et~al.}(2020){Kirsten}, {Snelders}, {Jenkins}, {Nimmo},
  {den van Eijnden}, {Hessels}, {Gawronski}, \& {Yang}}]{kirsten2020}
{Kirsten}, F., {Snelders}, M., {Jenkins}, M., {et~al.} 2020, arXiv e-prints,
  arXiv:2007.05101.
\newblock \doarXiv{2007.05101}

\bibitem[{{Kokubo} {et~al.}(2017){Kokubo}, {Mitsuda}, {Sugai}, {Ozaki},
  {Minowa}, {Hattori}, {Hayano}, {Matsubayashi}, {Shimono}, {Sako}, \&
  {Doi}}]{kokubo2017}
{Kokubo}, M., {Mitsuda}, K., {Sugai}, H., {et~al.} 2017, \apj, 844, 95,
  \dodoi{10.3847/1538-4357/aa7b2d}

\bibitem[{{Levin} {et~al.}(2020){Levin}, {Beloborodov}, \&
  {Bransgrove}}]{levin2020}
{Levin}, Y., {Beloborodov}, A.~M., \& {Bransgrove}, A. 2020, \apjl, 895, L30,
  \dodoi{10.3847/2041-8213/ab8c4c}

\bibitem[{{Li} {et~al.}(2020){Li}, {Lin}, {Xiong}, {Ge}, {Li}, {Li}, {Lu},
  {Zhang}, {Tuo}, {Nang}, {Zhang}, {Xiao}, {Chen}, {Song}, {Xu}, {Liu}, {Jia},
  {Cao}, {Zhang}, {Qu}, {Liao}, {Zhao}, {Tan}, {Nie}, {Zhao}, {Zheng}, {Zheng},
  {Luo}, {Cai}, {Li}, {Xue}, {Bu}, {Chang}, {Chen}, {Chen}, {Chen}, {Chen},
  {Chen}, {Cui}, {Cui}, {Deng}, {Dong}, {Du}, {Fu}, {Gao}, {Gao}, {Gao}, {Gu},
  {Guan}, {Guo}, {Han}, {Huang}, {Huo}, {Jiang}, {Jiang}, {Jin}, {Jin}, {Kong},
  {Li}, {Li}, {Li}, {Li}, {Li}, {Li}, {Li}, {Liang}, {Liu}, {Liu}, {Liu},
  {Liu}, {Liu}, {Lu}, {Lu}, {Luo}, {Ma}, {Meng}, {Ou}, {Sai}, {Shang}, {Song},
  {Sun}, {Tao}, {Wang}, {Wang}, {Wang}, {Wang}, {Wang}, {Wen}, {Wu}, {Wu},
  {Wu}, {Xiao}, {Yang}, {Yang}, {Yang}, {Yang}, {Yi}, {Yin}, {You}, {Zhang},
  {Zhang}, {Zhang}, {Zhang}, {Zhang}, {Zhang}, {Zhang}, {Zhang}, {Zhang},
  {Zhang}, {Zhang}, {Zhang}, {Zhang}, {Zhang}, {Zhang}, {Zhang}, {Zhou},
  {Zhou}, {Zhu}, {Zhu}, \& {Zhuang}}]{li2020}
{Li}, C.~K., {Lin}, L., {Xiong}, S.~L., {et~al.} 2020, arXiv e-prints,
  arXiv:2005.11071.
\newblock \doarXiv{2005.11071}

\bibitem[{{Lorimer} {et~al.}(2007){Lorimer}, {Bailes}, {McLaughlin},
  {Narkevic}, \& {Crawford}}]{lbm+07}
{Lorimer}, D.~R., {Bailes}, M., {McLaughlin}, M.~A., {Narkevic}, D.~J., \&
  {Crawford}, F. 2007, Science, 318, 777.
\newblock \doarXiv{0709.4301}

\bibitem[{{Lu} {et~al.}(2020){Lu}, {Kumar}, \& {Zhang}}]{lu2020}
{Lu}, W., {Kumar}, P., \& {Zhang}, B. 2020, \mnras, 498, 1397,
  \dodoi{10.1093/mnras/staa2450}

\bibitem[{{Luo} {et~al.}(2020){Luo}, {Wang}, {Men}, {Zhang}, {Jiang}, {Xu},
  {Wang}, {Lee}, {Han}, {Zhang}, {Caballero}, {Chen}, {Chen}, {Gan}, {Guo},
  {Hao}, {Huang}, {Jiang}, {Li}, {Li}, {Li}, {Luo}, {Pan}, {Pei}, {Qian},
  {Sun}, {Wang}, {Wang}, {Wen}, {Xu}, {Xu}, {Yan}, {Yan}, {Yu}, {Yuan},
  {Zhang}, \& {Zhu}}]{luo2020}
{Luo}, R., {Wang}, B.~J., {Men}, Y.~P., {et~al.} 2020, \nat, 586, 693–696,
  \dodoi{10.1038/s41586-020-2827-2}

\bibitem[{{Lyutikov} {et~al.}(2020){Lyutikov}, {Barkov}, \&
  {Giannios}}]{lyutikov2020}
{Lyutikov}, M., {Barkov}, M.~V., \& {Giannios}, D. 2020, \apjl, 893, L39,
  \dodoi{10.3847/2041-8213/ab87a4}

\bibitem[{{Macquart} {et~al.}(2020{\natexlab{a}}){Macquart}, {Prochaska},
  {McQuinn}, {Bannister}, {Bhandari}, {Day}, {Deller}, {Ekers}, {James},
  {Marnoch}, {Os{\l}owski}, {Phillips}, {Ryder}, {Scott}, {Shannon}, \&
  {Tejos}}]{mpm+20}
{Macquart}, J.~P., {Prochaska}, J.~X., {McQuinn}, M., {et~al.}
  2020{\natexlab{a}}, \nat, 581, 391, \dodoi{10.1038/s41586-020-2300-2}

\bibitem[{{Macquart} {et~al.}(2020{\natexlab{b}}){Macquart}, {Prochaska},
  {McQuinn}, {Bannister}, {Bhandari}, {Day}, {Deller}, {Ekers}, {James},
  {Marnoch}, {Os{\l}owski}, {Phillips}, {Ryder}, {Scott}, {Shannon}, \&
  {Tejos}}]{macquart2020}
---. 2020{\natexlab{b}}, \nat, 581, 391, \dodoi{10.1038/s41586-020-2300-2}

\bibitem[{{Mannings} {et~al.}(2020){Mannings}, {Fong}, {Simha}, {Prochaska},
  {Rafelski}, {Kilpatrick}, {Tejos}, {Heintz}, {Bhandari}, {Day}, {Deller},
  {Ryder}, {Shannon}, \& {Tendulkar}}]{mannings2020}
{Mannings}, A.~G., {Fong}, W.-f., {Simha}, S., {et~al.} 2020, arXiv e-prints,
  arXiv:2012.11617.
\newblock \doarXiv{2012.11617}

\bibitem[{{Marcote} {et~al.}(2017){Marcote}, {Paragi}, {Hessels}, {Keimpema},
  {van Langevelde}, {Huang}, {Bassa}, {Bogdanov}, {Bower}, {Burke-Spolaor},
  {Butler}, {Campbell}, {Chatterjee}, {Cordes}, {Demorest}, {Garrett}, {Ghosh},
  {Kaspi}, {Law}, {Lazio}, {McLaughlin}, {Ransom}, {Salter}, {Scholz},
  {Seymour}, {Siemion}, {Spitler}, {Tendulkar}, \& {Wharton}}]{mph+17}
{Marcote}, B., {Paragi}, Z., {Hessels}, J.~W.~T., {et~al.} 2017, \apjl, 834,
  L8.
\newblock \doarXiv{1701.01099}

\bibitem[{{Marcote} {et~al.}(2020){Marcote}, {Nimmo}, {Hessels}, {Tendulkar},
  {Bassa}, {Paragi}, {Keimpema}, {Bhardwaj}, {Karuppusamy}, {Kaspi}, {Law},
  {Michilli}, {Aggarwal}, {Andersen}, {Archibald}, {Bandura}, {Bower}, {Boyle},
  {Brar}, {Burke-Spolaor}, {Butler}, {Cassanelli}, {Chawla}, {Demorest},
  {Dobbs}, {Fonseca}, {Giri}, {Good}, {Gourdji}, {Josephy}, {Kirichenko},
  {Kirsten}, {Landecker}, {Lang}, {Lazio}, {Li}, {Lin}, {Linford}, {Masui},
  {Mena-Parra}, {Naidu}, {Ng}, {Patel}, {Pen}, {Pleunis}, {Rafiei-Ravandi},
  {Rahman}, {Renard}, {Scholz}, {Siegel}, {Smith}, {Stairs}, {Vanderlinde}, \&
  {Zwaniga}}]{mnh+20}
{Marcote}, B., {Nimmo}, K., {Hessels}, J.~W.~T., {et~al.} 2020, \nat, 577, 190,
  \dodoi{10.1038/s41586-019-1866-z}

\bibitem[{{Margalit} \& {Metzger}(2018)}]{mm18}
{Margalit}, B., \& {Metzger}, B.~D. 2018, \apjl, 868, L4,
  \dodoi{10.3847/2041-8213/aaedad}

\bibitem[{{Marino} {et~al.}(2013){Marino}, {Rosales-Ortega}, {S{\'a}nchez},
  {Gil de Paz}, {V{\'\i}lchez}, {Miralles-Caballero}, {Kehrig},
  {P{\'e}rez-Montero}, {Stanishev}, {Iglesias-P{\'a}ramo}, {D{\'\i}az},
  {Castillo-Morales}, {Kennicutt}, {L{\'o}pez-S{\'a}nchez}, {Galbany},
  {Garc{\'\i}a-Benito}, {Mast}, {Mendez-Abreu}, {Monreal-Ibero}, {Husemann},
  {Walcher}, {Garc{\'\i}a-Lorenzo}, {Masegosa}, {Del Olmo Orozco},
  {Mour{\~a}o}, {Ziegler}, {Moll{\'a}}, {Papaderos},
  {S{\'a}nchez-Bl{\'a}zquez}, {Gonz{\'a}lez Delgado}, {Falc{\'o}n-Barroso},
  {Roth}, {van de Ven}, \& {Califa Team}}]{marino2013}
{Marino}, R.~A., {Rosales-Ortega}, F.~F., {S{\'a}nchez}, S.~F., {et~al.} 2013,
  \aap, 559, A114, \dodoi{10.1051/0004-6361/201321956}

\bibitem[{{Martins} {et~al.}(2005){Martins}, {Schaerer}, \&
  {Hillier}}]{martins2005}
{Martins}, F., {Schaerer}, D., \& {Hillier}, D.~J. 2005, \aap, 436, 1049,
  \dodoi{10.1051/0004-6361:20042386}

\bibitem[{{Mereghetti} {et~al.}(2020){Mereghetti}, {Savchenko}, {Ferrigno},
  {G{\"o}tz}, {Rigoselli}, {Tiengo}, {Bazzano}, {Bozzo}, {Coleiro},
  {Courvoisier}, {Doyle}, {Goldwurm}, {Hanlon}, {Jourdain}, {von Kienlin},
  {Lutovinov}, {Martin-Carrillo}, {Molkov}, {Natalucci}, {Onori}, {Panessa},
  {Rodi}, {Rodriguez}, {S{\'a}nchez-Fern{\'a}ndez}, {Sunyaev}, \&
  {Ubertini}}]{mereghetti2020}
{Mereghetti}, S., {Savchenko}, V., {Ferrigno}, C., {et~al.} 2020, \apjl, 898,
  L29, \dodoi{10.3847/2041-8213/aba2cf}

\bibitem[{{Michilli} {et~al.}(2018){Michilli}, {Seymour}, {Hessels}, {Spitler},
  {Gajjar}, {Archibald}, {Bower}, {Chatterjee}, {Cordes}, {Gourdji}, {Heald},
  {Kaspi}, {Law}, {Sobey}, {Adams}, {Bassa}, {Bogdanov}, {Brinkman},
  {Demorest}, {Fernandez}, {Hellbourg}, {Lazio}, {Lynch}, {Maddox}, {Marcote},
  {McLaughlin}, {Paragi}, {Ransom}, {Scholz}, {Siemion}, {Tendulkar}, {Van
  Rooy}, {Wharton}, \& {Whitlow}}]{msh+18}
{Michilli}, D., {Seymour}, A., {Hessels}, J. W.~T., {et~al.} 2018, Nature, 533,
  132

\bibitem[{{Mottez} {et~al.}(2020){Mottez}, {Voisin}, \& {Zarka}}]{mvz20}
{Mottez}, F., {Voisin}, G., \& {Zarka}, P. 2020, arXiv e-prints,
  arXiv:2002.12834.
\newblock \doarXiv{2002.12834}

\bibitem[{Newville {et~al.}(2014)Newville, Stensitzki, Allen, \&
  Ingargiola}]{lmfit}
Newville, M., Stensitzki, T., Allen, D.~B., \& Ingargiola, A. 2014, {LMFIT:
  Non-Linear Least-Square Minimization and Curve-Fitting for Python}, 0.8.0,
  Zenodo, \dodoi{10.5281/zenodo.11813}

\bibitem[{{Nimmo} {et~al.}(2020){Nimmo}, {Hessels}, {Keimpema}, {Archibald},
  {Cordes}, {Karuppusamy}, {Kirsten}, {Li}, {Marcote}, \& {Paragi}}]{nimmo2020}
{Nimmo}, K., {Hessels}, J.~W.~T., {Keimpema}, A., {et~al.} 2020, arXiv
  e-prints, arXiv:2010.05800.
\newblock \doarXiv{2010.05800}

\bibitem[{{Olausen} \& {Kaspi}(2014)}]{ok14}
{Olausen}, S.~A., \& {Kaspi}, V.~M. 2014, \apjs, 212, 6,
  \dodoi{10.1088/0067-0049/212/1/6}

\bibitem[{Osterbrock \& Ferland(2006)}]{osterbrock2006astrophysics}
Osterbrock, D.~E., \& Ferland, G.~J. 2006, Astrophysics Of Gas Nebulae and
  Active Galactic Nuclei (University science books)

\bibitem[{{Pascual} {et~al.}(2019){Pascual}, {Cardiel}, {Gil de Paz},
  {Carasco}, {Gallego}, {Iglesias-P{\'a}ramo}, \& {Cedazo}}]{pascual2019}
{Pascual}, S., {Cardiel}, N., {Gil de Paz}, A., {et~al.} 2019, in Highlights on
  Spanish Astrophysics X, ed. B.~{Montesinos}, A.~{Asensio Ramos},
  F.~{Buitrago}, R.~{Sch{\"o}del}, E.~{Villaver}, S.~{P{\'e}rez-Hoyos}, \&
  I.~{Ord{\'o}{\~n}ez-Etxeberria}, 227--227

\bibitem[{{Pascual} {et~al.}(2018){Pascual}, {Cardiel}, {Picazo-Sanchez},
  {Castillo-Morales}, \& {Gil de Paz}}]{pascual2018}
{Pascual}, S., {Cardiel}, N., {Picazo-Sanchez}, P., {Castillo-Morales}, A., \&
  {Gil de Paz}, A. 2018, {guaix-ucm/megaradrp: v0.8}, v0.8,  Zenodo,
  \dodoi{10.5281/zenodo.2206856}

\bibitem[{{Pecaut} \& {Mamajek}(2013)}]{pecaut2013}
{Pecaut}, M.~J., \& {Mamajek}, E.~E. 2013, \apjs, 208, 9,
  \dodoi{10.1088/0067-0049/208/1/9}

\bibitem[{{Petroff} {et~al.}(2019){Petroff}, {Hessels}, \& {Lorimer}}]{phl19}
{Petroff}, E., {Hessels}, J.~W.~T., \& {Lorimer}, D.~R. 2019, \aapr, 27, 4,
  \dodoi{10.1007/s00159-019-0116-6}

\bibitem[{{Platts} {et~al.}(2018){Platts}, {Weltman}, {Walters}, {Tendulkar},
  {Gordin}, \& {Kandhai}}]{pww+18}
{Platts}, E., {Weltman}, A., {Walters}, A., {et~al.} 2018, arXiv e-prints.
\newblock \doarXiv{1810.05836}

\bibitem[{{Pleunis} {et~al.}(2020){Pleunis}, {Michilli}, {Bassa}, {Hessels},
  {Naidu}, {Andersen}, {Chawla}, {Fonseca}, {Gopinath}, {Kaspi}, {Kondratiev},
  {Li}, {Bhardwaj}, {Boyle}, {Brar}, {Cassanelli}, {Gupta}, {Josephy},
  {Karuppusamy}, {Keimpema}, {Kirsten}, {Leung}, {Marcote}, {Masui},
  {Mckinven}, {Meyers}, {Ng}, {Nimmo}, {Paragi}, {Rahman}, {Scholz}, {Shin},
  {Smith}, {Stairs}, \& {Tendulkar}}]{pleunis2020}
{Pleunis}, Z., {Michilli}, D., {Bassa}, C.~G., {et~al.} 2020, arXiv e-prints,
  arXiv:2012.08372.
\newblock \doarXiv{2012.08372}

\bibitem[{{Popov}(2016)}]{popov2016}
{Popov}, S.~B. 2016, Astronomical and Astrophysical Transactions, 29, 183.
\newblock \doarXiv{1507.08192}

\bibitem[{{Popov}(2020)}]{popov2020}
---. 2020, Research Notes of the American Astronomical Society, 4, 98,
  \dodoi{10.3847/2515-5172/aba0af}

\bibitem[{{Prochaska} {et~al.}(2019){Prochaska}, {Macquart}, {McQuinn},
  {Simha}, {Shannon}, {Day}, {Marnoch}, {Ryder}, {Deller}, {Bannister},
  {Bhandari}, {Bordoloi}, {Bunton}, {Cho}, {Flynn}, {Mahony}, {Phillips},
  {Qiu}, \& {Tejos}}]{pmm+19}
{Prochaska}, J.~X., {Macquart}, J.-P., {McQuinn}, M., {et~al.} 2019, Science,
  366, 231, \dodoi{10.1126/science.aay0073}

\bibitem[{{Rajwade} {et~al.}(2020){Rajwade}, {Mickaliger}, {Stappers},
  {Morello}, {Agarwal}, {Bassa}, {Breton}, {Caleb}, {Karastergiou}, {Keane}, \&
  {Lorimer}}]{rms+20}
{Rajwade}, K.~M., {Mickaliger}, M.~B., {Stappers}, B.~W., {et~al.} 2020,
  \mnras, 495, 3551, \dodoi{10.1093/mnras/staa1237}

\bibitem[{{Ravi} {et~al.}(2019){Ravi}, {Catha}, {D'Addario}, {Djorgovski},
  {Hallinan}, {Hobbs}, {Kocz}, {Kulkarni}, {Shi}, {Vedantham}, {Weinreb}, \&
  {Woody}}]{rcd+19}
{Ravi}, V., {Catha}, M., {D'Addario}, L., {et~al.} 2019, \nat, 572, 352,
  \dodoi{10.1038/s41586-019-1389-7}

\bibitem[{{Reig}(2011)}]{reig2011}
{Reig}, P. 2011, \apss, 332, 1, \dodoi{10.1007/s10509-010-0575-8}

\bibitem[{{Ridnaia} {et~al.}(2020){Ridnaia}, {Svinkin}, {Frederiks}, {Bykov},
  {Popov}, {Aptekar}, {Golenetskii}, {Lysenko}, {Tsvetkova}, {Ulanov}, \&
  {Cline}}]{ridnaia2020}
{Ridnaia}, A., {Svinkin}, D., {Frederiks}, D., {et~al.} 2020, arXiv e-prints,
  arXiv:2005.11178.
\newblock \doarXiv{2005.11178}

\bibitem[{{Safarzadeh} {et~al.}(2020){Safarzadeh}, {Prochaska}, {Heintz}, \&
  {Fong}}]{safarzadeh2020}
{Safarzadeh}, M., {Prochaska}, J.~X., {Heintz}, K.~E., \& {Fong}, W.-f. 2020,
  arXiv e-prints, arXiv:2009.11735.
\newblock \doarXiv{2009.11735}

\bibitem[{{Schlafly} \& {Finkbeiner}(2011)}]{schlafly2011}
{Schlafly}, E.~F., \& {Finkbeiner}, D.~P. 2011, \apj, 737, 103,
  \dodoi{10.1088/0004-637X/737/2/103}

\bibitem[{{Scholz} {et~al.}(2017){Scholz}, {Bogdanov}, {Hessels}, {Lynch},
  {Spitler}, {Bassa}, {Bower}, {Burke-Spolaor}, {Butler}, {Chatterjee},
  {Cordes}, {Gourdji}, {Kaspi}, {Law}, {Marcote}, {McLaughlin}, {Michilli},
  {Paragi}, {Ransom}, {Seymour}, {Tendulkar}, \& {Wharton}}]{sbh+17}
{Scholz}, P., {Bogdanov}, S., {Hessels}, J.~W.~T., {et~al.} 2017, \apj, 846,
  80, \dodoi{10.3847/1538-4357/aa8456}

\bibitem[{{Scholz} {et~al.}(2020){Scholz}, {Cook}, {Cruces}, {Hessels},
  {Kaspi}, {Majid}, {Naidu}, {Pearlman}, {Spitler}, {Bandura}, {Bhardwaj},
  {Cassanelli}, {Chawla}, {Gaensler}, {Good}, {Josephy}, {Karuppusamy},
  {Keimpema}, {Kirichenko}, {Kirsten}, {Kocz}, {Leung}, {Marcote}, {Masui},
  {Mena-Parra}, {Merryfield}, {Michilli}, {Naudet}, {Nimmo}, {Pleunis},
  {Prince}, {Rafiei-Ravandi}, {Rahman}, {Shin}, {Smith}, {Stairs}, {Tendulkar},
  \& {Vanderlinde}}]{scholz2020}
{Scholz}, P., {Cook}, A., {Cruces}, M., {et~al.} 2020, \apj, 901, 165,
  \dodoi{10.3847/1538-4357/abb1a8}

\bibitem[{{Shannon} {et~al.}(2018){Shannon}, {Macquart}, {Bannister}, {Ekers},
  {James}, {Os{\l}owski}, {Qiu}, {Sammons}, {Hotan}, {Voronkov}, {Beresford},
  {Brothers}, {Brown}, {Bunton}, {Chippendale}, {Haskins}, {Leach},
  {Marquarding}, {McConnell}, {Pilawa}, {Sadler}, {Troup}, {Tuthill},
  {Whiting}, {Allison}, {Anderson}, {Bell}, {Collier}, {G{\"u}rkan}, {Heald},
  \& {Riseley}}]{smb+18}
{Shannon}, R.~M., {Macquart}, J.-P., {Bannister}, K.~W., {et~al.} 2018, \nat,
  562, 386, \dodoi{10.1038/s41586-018-0588-y}

\bibitem[{{Sob'yanin}(2020)}]{sobyanin2020}
{Sob'yanin}, D.~N. 2020, \mnras, 497, 1001, \dodoi{10.1093/mnras/staa1976}

\bibitem[{{Spitler} {et~al.}(2014){Spitler}, {Cordes}, {Hessels}, {Lorimer},
  {McLaughlin}, {Chatterjee}, {Crawford}, {Deneva}, {Kaspi}, {Wharton},
  {Allen}, {Bogdanov}, {Brazier}, {Camilo}, {Freire}, {Jenet},
  {Karako-Argaman}, {Knispel}, {Lazarus}, {Lee}, {van Leeuwen}, {Lynch},
  {Ransom}, {Scholz}, {Siemens}, {Stairs}, {Stovall}, {Swiggum},
  {Venkataraman}, {Zhu}, {Aulbert}, \& {Fehrmann}}]{sch+14}
{Spitler}, L.~G., {Cordes}, J.~M., {Hessels}, J.~W.~T., {et~al.} 2014, \apj,
  790, 101.
\newblock \doarXiv{1404.2934}

\bibitem[{{Spitler} {et~al.}(2016){Spitler}, {Scholz}, {Hessels}, {Bogdanov},
  {Brazier}, {Camilo}, {Chatterjee}, {Cordes}, {Crawford}, {Deneva}, {Ferdman},
  {Freire}, {Kaspi}, {Lazarus}, {Lynch}, {Madsen}, {McLaughlin}, {Patel},
  {Ransom}, {Seymour}, {Stairs}, {Stappers}, {van Leeuwen}, \& {Zhu}}]{ssh+16a}
{Spitler}, L.~G., {Scholz}, P., {Hessels}, J.~W.~T., {et~al.} 2016, \nat, 531,
  202.
\newblock \doarXiv{1603.00581}

\bibitem[{{Tavani} {et~al.}(2020){Tavani}, {Casentini}, {Ursi}, {Verrecchia},
  {Addis}, {Antonelli}, {Argan}, {Barbiellini}, {Baroncelli}, {Bernardi},
  {Bianchi}, {Bulgarelli}, {Caraveo}, {Cardillo}, {Cattaneo}, {Chen}, {Costa},
  {Del Monte}, {Di Cocco}, {Di Persio}, {Donnarumma}, {Evangelista}, {Feroci},
  {Ferrari}, {Fioretti}, {Fuschino}, {Galli}, {Gianotti}, {Giuliani},
  {Labanti}, {Lazzarotto}, {Lipari}, {Longo}, {Lucarelli}, {Magro},
  {Marisaldi}, {Mereghetti}, {Morelli}, {Morselli}, {Naldi}, {Pacciani},
  {Parmiggiani}, {Paoletti}, {Pellizzoni}, {Perri}, {Perotti}, {Piano},
  {Picozza}, {Pilia}, {Pittori}, {Puccetti}, {Pupillo}, {Rapisarda},
  {Rappoldi}, {Rubini}, {Setti}, {Soffitta}, {Trifoglio}, {Trois},
  {Vercellone}, {Vittorini}, {Giommi}, \& {D' Amico}}]{tavani2020}
{Tavani}, M., {Casentini}, C., {Ursi}, A., {et~al.} 2020, arXiv e-prints,
  arXiv:2005.12164.
\newblock \doarXiv{2005.12164}

\bibitem[{{Tendulkar}(2014)}]{tendulkar2014}
{Tendulkar}, S.~P. 2014, PhD thesis, California Institute of Technology

\bibitem[{{Tendulkar} {et~al.}(2016){Tendulkar}, {Kaspi}, \& {Patel}}]{tkp16}
{Tendulkar}, S.~P., {Kaspi}, V.~M., \& {Patel}, C. 2016, \apj, 827, 59,
  \dodoi{10.3847/0004-637X/827/1/59}

\bibitem[{{Tendulkar} {et~al.}(2017){Tendulkar}, {Bassa}, {Cordes}, {Bower},
  {Law}, {Chatterjee}, {Adams}, {Bogdanov}, {Burke-Spolaor}, {Butler},
  {Demorest}, {Hessels}, {Kaspi}, {Lazio}, {Maddox}, {Marcote}, {McLaughlin},
  {Paragi}, {Ransom}, {Scholz}, {Seymour}, {Spitler}, {van Langevelde}, \&
  {Wharton}}]{tbc+17}
{Tendulkar}, S.~P., {Bassa}, C.~G., {Cordes}, J.~M., {et~al.} 2017, \apjl, 834,
  L7.
\newblock \doarXiv{1701.01100}

\bibitem[{{Torres} {et~al.}(2012){Torres}, {Rea}, {Esposito}, {Li}, {Chen}, \&
  {Zhang}}]{torres2012}
{Torres}, D.~F., {Rea}, N., {Esposito}, P., {et~al.} 2012, \apj, 744, 106,
  \dodoi{10.1088/0004-637X/744/2/106}

\bibitem[{{Tsygankov} {et~al.}(2016){Tsygankov}, {Mushtukov}, {Suleimanov}, \&
  {Poutanen}}]{tsygankov2016}
{Tsygankov}, S.~S., {Mushtukov}, A.~A., {Suleimanov}, V.~F., \& {Poutanen}, J.
  2016, \mnras, 457, 1101, \dodoi{10.1093/mnras/stw046}

\bibitem[{Virtanen {et~al.}(2020)Virtanen, Gommers, Oliphant, Haberland, Reddy,
  Cournapeau, Burovski, Peterson, Weckesser, Bright, {van der Walt}, Brett,
  Wilson, Millman, Mayorov, Nelson, Jones, Kern, Larson, Carey, Polat, Feng,
  Moore, {VanderPlas}, Laxalde, Perktold, Cimrman, Henriksen, Quintero, Harris,
  Archibald, Ribeiro, Pedregosa, {van Mulbregt}, \& {SciPy 1.0
  Contributors}}]{2020SciPy-NMeth}
Virtanen, P., Gommers, R., Oliphant, T.~E., {et~al.} 2020, Nature Methods, 17,
  261, \dodoi{10.1038/s41592-019-0686-2}

\bibitem[{{Walter} {et~al.}(2015){Walter}, {Lutovinov}, {Bozzo}, \&
  {Tsygankov}}]{walter2015}
{Walter}, R., {Lutovinov}, A.~A., {Bozzo}, E., \& {Tsygankov}, S.~S. 2015,
  \aapr, 23, 2, \dodoi{10.1007/s00159-015-0082-6}

\bibitem[{{Weng*} {et~al.}(2021){Weng*}, {Pan*}, {Qian*}, {Jiang}, {Ge}, {Yan},
  \& {Liu}}]{lsi61303_atel}
{Weng*}, S.-S., {Pan*}, Z., {Qian*}, L., {et~al.} 2021, The Astronomer's
  Telegram, 14297, 1

\bibitem[{{Worthey} \& {Lee}(2011)}]{worthey2011}
{Worthey}, G., \& {Lee}, H.-c. 2011, \apjs, 193, 1,
  \dodoi{10.1088/0067-0049/193/1/1}

\bibitem[{{Yang} \& {Zou}(2020{\natexlab{a}})}]{yangzou2020}
{Yang}, H., \& {Zou}, Y.-C. 2020{\natexlab{a}}, \apjl, 893, L31,
  \dodoi{10.3847/2041-8213/ab800f}

\bibitem[{{Yang} \& {Zou}(2020{\natexlab{b}})}]{yang2020}
---. 2020{\natexlab{b}}, \apjl, 893, L31, \dodoi{10.3847/2041-8213/ab800f}

\bibitem[{{Zanazzi} \& {Lai}(2020)}]{zanazzi2020}
{Zanazzi}, J.~J., \& {Lai}, D. 2020, \apjl, 892, L15,
  \dodoi{10.3847/2041-8213/ab7cdd}

\bibitem[{{Zhang} \& {Gao}(2020)}]{zhanggao2020}
{Zhang}, X., \& {Gao}, H. 2020, \mnras, 498, L1, \dodoi{10.1093/mnrasl/slaa116}

\bibitem[{{Zinnecker} \& {Yorke}(2007)}]{zinnecker2007}
{Zinnecker}, H., \& {Yorke}, H.~W. 2007, \araa, 45, 481,
  \dodoi{10.1146/annurev.astro.44.051905.092549}

\end{thebibliography}
\bibliographystyle{aasjournal}



\end{document}